\documentclass[twocolumn]{aastex62}

\usepackage{ amssymb }

\usepackage{ skull }
\usepackage{natbib}

\usepackage{lineno}
\newcommand{\cii}{[CII]$_{158\mu m}\,$}

\graphicspath{{./}{figures/}}

\received{XXXX}
\accepted{YYYY}

\shorttitle{}
\shortauthors{}

\begin{document}

\title{Significant Dust-Obscured Star Formation in Luminous Lyman-break Galaxies at $z\sim7$-8}

\author[0000-0001-9746-0924]{Sander Schouws}
\affil{Leiden Observatory, Leiden University, NL-2300 RA Leiden, Netherlands}

\author[0000-0001-7768-5309]{Mauro Stefanon}
\affil{Leiden Observatory, Leiden University, NL-2300 RA Leiden, Netherlands}

\author[0000-0002-4989-2471]{Rychard Bouwens}
\affil{Leiden Observatory, Leiden University, NL-2300 RA Leiden, Netherlands}

\author[0000-0001-7768-5309]{Renske Smit}
\affil{Cavendish Laboratory, University of Cambridge, 19 JJ Thomson Avenue, Cambridge CB3 0HE, UK}

\author[0000-0001-6586-8845]{Jacqueline Hodge}
\affil{Leiden Observatory, Leiden University, NL-2300 RA Leiden, Netherlands}

\author[0000-0002-2057-5376]{Ivo Labb\'e}
\affil{Centre for Astrophysics and SuperComputing, Swinburne, University of Technology, Hawthorn, Victoria, 3122, Australia}


\author[0000-0002-4205-9567]{Hiddo Algera}
\affil{Leiden Observatory, Leiden University, NL-2300 RA Leiden, Netherlands}

\author[0000-0002-3952-8588]{Leindert Boogaard}
\affil{Leiden Observatory, Leiden University, NL-2300 RA Leiden, Netherlands}

\author[0000-0002-6719-380X]{Stefano Carniani}
\affil{Scuola Normale Superiore, Piazza dei Cavalieri 7, I-56126 Pisa, Italy}

\author[0000-0001-7440-8832]{Yoshinobu Fudamoto}
\affil{Research Institute for Science and Engineering, Waseda University, 3-4-1 Okubo, Shinjuku, Tokyo 169-8555, Japan}
\affil{National Astronomical Observatory of Japan, 2-21-1, Osawa, Mitaka, Tokyo, Japan}

\author[0000-0002-4884-6756]{Benne W. Holwerda}
\affil{Department of Physics and Astronomy, University of Louisville, Natural Science 102, Louisville, KY 40292, USA}

\author[0000-0002-8096-2837]{Garth D. Illingworth}
\affil{UCO/Lick Observatory, University of California, Santa Cruz, 1156 High St, Santa Cruz, CA 95064, USA}

\author[0000-0002-4985-3819]{Roberto Maiolino}
\affil{Cavendish Laboratory, University of Cambridge, 19 J. J. Thomson Avenue, Cambridge CB3 0HE, UK}
\affil{Kavli Institute for Cosmology, University of Cambridge, Madingley Road, Cambridge CB3 0HA, UK}
\affil{Department of Physics and Astronomy, University College London, Gower Street, London WC1E 6BT, UK}

\author[0000-0003-0695-4414]{Michael Maseda}
\affil{Leiden Observatory, Leiden University, NL-2300 RA Leiden, Netherlands}

\author[0000-0001-5851-6649]{Pascal Oesch}
\affil{Departement d’Astronomie, Universit\'ede Gen\'eeve, 51 Ch. des Maillettes, CH-1290 Versoix, Switzerland}
\affil{International Associate, Cosmic Dawn Center (DAWN), Niels Bohr Institute, University of Copenhagen and DTU-Space, Technical University of Denmark}

\author[0000-0002-4389-832X]{Paul van der Werf}
\affil{Leiden Observatory, Leiden University, NL-2300 RA Leiden, Netherlands}



\begin{abstract}
We make use of ALMA continuum observations of 15 luminous Lyman-break galaxies at $z\sim7$-8 to probe their dust-obscured star-formation.  These observations are 
sensitive enough to probe to obscured SFRs of 20 $M_{\odot}$/yr ($3\sigma$).  Six of the targeted galaxies show significant ($\geq$3$\sigma$) dust continuum detections, more than doubling the number of known dust-detected galaxies at $z>6.5$.  Their IR luminosities range from $2.7\times10^{11}$ $L_{\odot}$ to $1.1\times10^{12}$ $L_{\odot}$, equivalent to obscured SFRs of 20 to 105 $M_{\odot}$/yr.  We use our results to quantify the correlation of the infrared excess IRX on the UV-continuum slope $\beta_{UV}$ and stellar mass.  Our results are most consistent with an SMC attenuation curve for intrinsic $UV$-slopes $\beta_{UV,intr}$ of $-2.63$ and most consistent with an attenuation curve in-between SMC and Calzetti for $\beta_{UV,intr}$ slopes of $-2.23$, assuming a dust temperature $T_d$ of 50 K.   Our fiducial IRX-stellar mass results at $z\sim7$-8 are consistent with marginal evolution from $z\sim0$.
We then show how both results depend on $T_d$.  For our six dust-detected sources, we estimate their dust masses and find that they are consistent with dust production from SNe if the dust destruction is low ($<$90\%). Finally we determine the contribution of dust-obscured star formation to the star formation rate density for $UV$ luminous ($<$$-$21.5 mag: $\gtrsim$1.7$L_{UV} ^*$) $z\sim7$-8 galaxies, finding that the total SFR density at $z\sim7$ and $z\sim8$ from bright galaxies is 0.18$_{-0.10}^{+0.08}$ dex and 0.20$_{-0.09}^{+0.05}$ dex higher, respectively, i.e. $\sim$$\frac{1}{3}$ of the star formation in $\gtrsim$1.7$L_{UV} ^*$ galaxies at $z\sim7$-8 is obscured by dust.
\end{abstract} 



\section{Introduction} \label{sec:intro}

One major uncertainty in our understanding of galaxy formation during the Reionization Epoch regards the role of early dust build-up.  Deep surveys in the rest-$UV$ with both ground and space-based facilities have enabled the identification of large samples of galaxies at $z>2$ and measurement of their \textit{UV}-based star formation rates up to redshift $z\sim$11 \citep[e.g., ][]{Bouwens_2011, Bouwens_2014, Bouwens_2015, Bouwens_2016, Bouwens_2017, Ellis_2012, Laporte_2012, mclure2013, Schenker_2013, Finkelstein_2015, Kawamata_2018, Oesch_2014,Oesch_2015,Oesch_2018, mcleod_2016}.  The rest-$UV$ view only accounts for unobscured star formation and could be significantly impacted by the presence of dust in galaxies.  Fortunately, accurate measurements of the dust-obscured star formation rates in galaxies to $z<3$ have been available for the past ten years, thanks to deep and wide-field far-infrared (FIR) observations from Herschel and Spitzer \citep[][]{Reddy_2006, Daddi_2007, magnelli2009, magnelli2013, Magnelli_2020, Karim_2011, Cucciati2012, schreiber2015A&A...575A..74S, alvarez-marquez2016}.  

Less is known however about star formation in $z>3$ galaxies that might be obscured by dust \citep[e.g.,][]{J_Bouwens_2016_dust,dunlop2016, Casey_2018, wang2019W, Williams_2019}.  Because of the limited sensitivity and increasing source confusion suffered by Herschel and Spitzer, the obscured star formation in $z>3$ galaxies could initially only be studied in the most luminous systems \citep[e.g.][]{riechers2013Natur.496..329R,Oteo_2016, marrone2018Natur.553...51M}.  Consequently, studies of the star formation rate density at $z>3$ often account for the impact of obscured star formation based on the $UV$ light alone.  Especially useful in this regard has been the empirical relation between the UV continuum slope $\beta_{UV}$ and the infrared excess (IRX=$L_{IR}/L_{UV}$).  In the local universe, the IRX has been found to show  a good correlation with $\beta_{UV}$ for star forming galaxies \citep{Meurer_1999,takeuchi2010A&A...514A...4T, Overzier_2010, Casey_2014}. Moreover, this same correlation seems to hold up to at least $z\sim3$ \citep{Reddy_2006,Reddy_2010,Reddy_2011,Daddi_2009,Overzier_2010,sklias2014A&A...561A.149S, Pannella_2015, J_Bouwens_2016_dust, alvarez-marquez2016, Whitaker_2017, Fudamoto201710.1093/mnras/stx1948, Fudamoto201910.1093/mnras/stz3248, Fudamoto2020arXiv200410760F, mclure2018_10.1093/mnras/sty522}.

The availability of new observations from the increasingly sensitive and high-resolution sub-millimeter facilities like the Atacama Large Millimeter/submillimeter Array (ALMA) and the NOrthern Extended Millimeter Array (NOEMA) have had an especially important impact on studies of obscured star formation and dust in $z>3$ galaxies \citep[e.g.][]{Hodge2020arXiv200400934H}.  Most importantly, these instruments have enabled the detection of obscured star formation in normal main sequence star forming galaxies at $z>$3 by utilizing deep field observations \citep{Aravena_2016,Aravena_2020,J_Bouwens_2016_dust,bouwens_2020, dunlop2016, mclure2013} and targeted studies \citep[e.g.][]{Capak2015_Natur.522..455C, Willott_2015, Knudsen_2017_10.1093/mnras/stw3066, Laporte_2017, Bowler_2018_10.1093/mnras/sty2368, Hashimoto_2019_10.1093/pasj/psz049, Tamura_2019, Fudamoto2020arXiv200410760F}, but the majority of observed high redshift sources remain undetected in the dust continuum \citep[e.g][]{Maiolino_2015_10.1093/mnras/stv1194,J_Bouwens_2016_dust,Carniani_2017,Carniani_2018,Smit_2018Natur.553..178S,Bowler_2018_10.1093/mnras/sty2368,Matthee_2017,Matthee_2019, Hashimoto_2018Natur.557..392H}.  This indicates that star formation in normal $z>3$ galaxies may be less obscured than in the local universe, especially in normal and small galaxies, and previous corrections to the UV derived star formation rates to determine the cosmic star formation history may be overestimated \citep[e.g.][]{J_Bouwens_2016_dust, Capak2015_Natur.522..455C, Willott_2015, Barisic_2017, Faisst_2017, Faisst_2020_10.1093/mnras/staa2545, Fudamoto201710.1093/mnras/stx1948, Fudamoto201910.1093/mnras/stz3248, Fudamoto_2020_NOT_ALPINE}.  \citet{Fudamoto2020arXiv200410760F}, in particular, find that the infrared excess of $z\sim4$-6 galaxies from ALPINE \citep{LeFevre_2020} can be much better explained with a steeper attenuation curve like SMC than with greyer attenuation curves like Calzetti.

Despite the significant amount of progress in detecting obscured star formation in galaxies to redshifts $z\sim6.5$, much less is known about the dust properties and obscured star formation rates of galaxies at $z>6.5$, with only 4 dust-continuum detections thus far identified in normal star-forming galaxies at $z\sim7$-8 \citep{Watson_2015_Natur.519..327W, Knudsen_2017_10.1093/mnras/stw3066, Laporte_2017,Bowler_2018_10.1093/mnras/sty2368,Hashimoto_2019_10.1093/pasj/psz049,Tamura_2019}.  Additionally, there have also been a modest number of prominent dust continuum detections in more extreme ULIRG-type (Ultra-Luminous InfraRed Galaxy) sources \citep{marrone2018Natur.553...51M} and QSOs at $z>6.5$ \citep{Venemans_2012,Venemans_2015b, Venemans_2015,Venemans_2017,Venemans_2018,Banados_2015,Mazzucchelli_2017}.

Further insight can be obtained into the dust properties of luminous $z>6$ galaxies by increasing the number of such galaxies targeted with sensitive dust-continuum observations.  In this spirit, we successfully proposed for ALMA continuum  observations on a significant sample of luminous $z\sim8$ galaxies to study the dust-continuum emission from these galaxies.  In addition, we also obtained sensitive ALMA-continuum observations on seven other bright $z\sim7$ galaxies as part of two other programs aimed at probing the \cii line.  Eight of these galaxies were identified as part of the \citet{Stefanon2017,Stefanon2019} study of bright star-forming galaxies at $z\gtrsim8$ in the COSMOS/UltraVISTA field, while the $z\sim7$ galaxies were primarily part of a similar selection of bright galaxies at $z\sim7$ (Schouws et al.\ 2021, \textit{in prep}).  Focus on the brightest and most massive galaxies in the early Universe is especially promising for the study of obscured star formation and dust, given the prevalence of dust in such galaxies at lower redshift \citep[e.g.,][]{Hodge2020arXiv200400934H}, and the observability of such sources with ALMA and NOEMA.

In this paper, we make use of these new ALMA continuum observations to characterize the dust properties of some of the most massive galaxies in the $z\sim7$-8 Universe.  Our new observations significantly add to the ALMA continuum observations currently available for galaxies at $z>6$, allowing us to quantify in much more detail both the dust emission and obscured star formation in these massive sources.  In addition, the new observations allow us to constrain the dust law for high-redshift sources more accurately and increase the sample of sources where we have estimates of dust masses, which is essential for studying the build-up of dust early after the Big Bang. 

The plan for this paper is as follows.  In  \textsection 2, we describe both the selection of our sources and the ALMA observations.  In \textsection 3, we present our new ALMA-continuum results and then combine these results with the available UV observations to derive the IR luminosities and infrared excesses for individual sources.  In \textsection 4, we use our new measurements to constrain the form of IRX-M$_{*}$ and IRX-$\beta$ relations, look at spatial offsets between $UV$ and dust-continuum light, and consider dust build-up in galaxies to $z\sim7$.  In \textsection 5, we briefly discuss our results, and finally, in \textsection 6, we summarize our conclusions.  Throughout this paper we assume a standard cosmology with $H_0=70$ km s$^{-1}$ Mpc$^{-1}$, $\Omega_m=0.3$ and $\Omega_{\lambda}=0.7$. Magnitudes are presented in the AB system \citep{oke_gunn_1983ApJ...266..713O}. For star formation rates and stellar masses we adopt a Chabrier IMF \citep{Chabrier_2003}.


\newcommand{\specialcell}[2][c]{%
  \begin{tabular}[#1]{@{}c@{}}#2\end{tabular}}

\begin{deluxetable*}{lcccccccc}
\tablecaption{Observational parameters of the ALMA observations\label{tab:tab1}}
\tablewidth{0pt}
\tablehead{\colhead{\specialcell[c]{Source \\ Name \\ }} & \colhead{\specialcell[c]{RA}} & \colhead{\specialcell[c]{DEC}}     & \colhead{\specialcell[c]{Beamwidth$^{a}$ \\ \\ (arcsec)}} & \colhead{\specialcell[c]{Integration\\time$^{b}$\\(min.)} } & \colhead{\specialcell[c]{Achieved \\ Sensitivity \\ ($\mu$Jy: 1$\sigma$)}} & \colhead{\specialcell[c]{Central\\Frequency\\(GHz)}} & \colhead{\specialcell[c]{Project\\Code}}}
\startdata
\multicolumn{8}{c}{$z\sim7$ Sample} \\
UVISTA-Z-001                                           & 10:00:43.36                                    & 02:37:51.3                                      & 1.47"$\times$1.21"                                   & 78.36                                                                & 17.9                                                 & 234.0                                                         & 2018.1.00085.S                                      \\
UVISTA-Z-007                                           & 09:58:46.21                                    & 02:28:45.8                                      & 1.40"$\times$1.19"                                   & 65.52                                                                & 17.4                                                 & 245.7                                                         & 2018.1.00085.S                                      \\
UVISTA-Z-009                                           & 10:01:52.30                                    & 02:25:42.3                                      & 1.38"$\times$1.20"                                   & 65.52                                                                & 19.0                                                 & 245.7                                                         & 2018.1.00085.S                                      \\
UVISTA-Z-010                                           & 10:00:28.12                                    & 01:47:54.5                                      & 1.44"$\times$1.18"                                   & 78.36                                                                & 14.7                                                 & 234.0                                                         & 2018.1.00085.S                                      \\
UVISTA-Z-013                                           & 09:59:19.35                                    & 02:46:41.3                                      & 1.45"$\times$1.18"                                   & 78.36                                                                & 15.0                                                 & 234.0                                                         & 2018.1.00085.S                                      \\
UVISTA-Z-019                                           & 10:00:29.89                                    & 01:46:46.4                                      & 1.39"$\times$1.18"                                   & 65.52                                                                & 18.0                                                 & 245.7                                                         & 2018.1.00085.S                                      \\
COS-3018                                                 & 10:00:30.19                                    & 02:15:59.8                                      & 1.63"$\times$1.49"$^{c}$                                   & 766.08                                                                & 7.0                                                   & 233.5                                                         & 2017.1.00604.S                                      \\
\multicolumn{8}{c}{$z\sim8$ Sample} \\
UVISTA-Y-001                                           & 09:57:47.90                                    & 02:20:43.7                                      & 1.17"$\times$0.96"                                   & 37.50                                                                & 13.4                                                   & 203.1                                                         & 2018.1.00236.S                                      \\
UVISTA-Y-002                                           & 10:02:12.56                                    & 02:30:45.7                                      & 1.17"$\times$0.96"                                   & 37.50                                                                & 13.3                                                   & 203.1                                                         & 2018.1.00236.S                                      \\
UVISTA-Y-003                                           & 10:00:32.32                                    & 01:44:31.3                                      & 0.60"$\times$0.52"                                   & 33.26                                                                & 19.1                                                   & 230.4                                                         & 2017.1.01217.S                                      \\
UVISTA-Y-004                                           & 10:00:58.49                                    & 01:49:56.0                                      & 1.03"$\times$0.94"                                   & 91.73                                                                & 9.3                                                    & 231.3                                                         & \specialcell[c]{2017.1.01217.S \& \\ 2018.A.00022.S}                                      \\
UVISTA-Y-005                                           & 10:00:31.89                                    & 01:57:50.2                                      & 1.17"$\times$0.96"                                   & 37.50                                                                & 12.9                                                   & 203.1                                                         & 2018.1.00236.S                                      \\
UVISTA-Y-006                                           & 10:00:12.51                                    & 02:03:00.5                                      & 1.17"$\times$0.96"                                   & 37.50                                                                & 13.2                                                   & 203.1                                                         & 2018.1.00236.S                                      \\
UVISTA-Y-007\tablenotemark{d}                                           & 09:59:02.57                                   & 02:38:06.1                                      & 0.60"$\times$0.53"                                   & 34.78                                                                & 17.9                                                   & 230.4                                                         & 2017.1.01217.S                                      \\
UVISTA-Y-009\tablenotemark{e}                                         & 09:59:09.62                                    & 02:45:09.7                                      & 0.60"$\times$0.53"                                   & 34.78                                                                & 17.9                                                   & 230.4                                                         & 2017.1.01217.S \\ \enddata
\tablecomments{
\textsuperscript{a}Beamsize as measured in the naturally weighted continuum images.
\textsuperscript{b}Corresponds to the total integration time in case of multiple tunings. In total, 25.7 hours of integration time on source is included in this analysis.
\textsuperscript{c}For this dataset we use a 1.5 arcsecond tapered continuum image, the full $\sim$0.25 arcsecond resolution data will be presented in Smit et al.\ (2021, \textit{in prep}).
\textsuperscript{d}Source name in ALMA observations is Y8.
\textsuperscript{e}Source name in ALMA observations is Y10.
\vspace{-0.6cm}}
\end{deluxetable*}

\section{Targets and ALMA Observations} \label{sec:selection}

The purpose of this Section is to provide a brief summary of the deep ALMA continuum observations and bright $z\sim7$-8 targets we analyze in this study as part of five separate ALMA programs.  A summary of the observations we describe in this Section is provided in Table \ref{tab:tab1}.

\subsection{ALMA Continuum Observations of Bright $z\sim7$ Galaxies}

As part of an effort to better characterize the physical properties of very luminous galaxies at $z\sim7$, we used an ALMA cycle-6 program (2018.1.00085.S: PI Schouws) to scan six luminous $z\sim7$ galaxies for the \cii line (Schouws et al.\ 2021, \textit{in prep}).  The six sources in the program were identified using the very deep optical, near-IR, and Spitzer/IRAC observations obtained over the 2 deg$^2$ COSMOS/UltraVISTA field from significant survey programs within COSMOS \citep{Scoville_2007, Capak_2007}, CFHT-LS \citep{Erben_2009A&A...493.1197E, Hildebrandt_2009A&A...498..725H}, UltraVISTA \citep{McCracken_2012A&A...544A.156M}, SPLASH \citep{Capak2013sptz.prop10042C}, and SMUVS \citep{Caputi_2017,Ashby_2018}.

Each of the sources was selected to show a strong break across the $Y$ and $z$ filters ($z - Y >$ 2 mag), show blue UV-continuum slopes, and show moderately blue or red Spitzer/IRAC [3.6]-[4.5] colors, i.e., $<-0.3$ or 
$\geq$0.3 (Schouws et al.\ 2021, \textit{in prep}). 
Significantly red or blue Spitzer/IRAC colors not only provide evidence for substantial [OIII]$_{5007\text{\AA}}$+H$\beta$ line emission in the sources, e.g., $\gtrsim$500\AA$\,$ 
\citep{Smit_2015} but also indicate which IRAC band the line emission contributes.  By combining this information with that available from the Lyman break, accurate redshift constraints can be derived $\Delta$$z$$\lesssim$0.15 
\citep[see also][]{Smit_2015, Roberts_Borsani_2016} on the six targeted galaxies, making it possible for us to efficiently scan for the 157.74$\mu$m [CII] line (hereafter [CII]$_{158\mu m}$).  A study of the \cii lines will be presented in Schouws et al.\ (2021, \textit{in prep}).  Here we focus on the dust-continuuum detections.  

All six targets were observed in Band 6 as part of ALMA program \#2018.1.00085.S using the most compact configuration for maximal sensitivity ($\sim$1.5 arcseconds resolution). The spectral setup was chosen to enable a scan for \cii  over essentially the full frequency range expected on the basis of our precise redshift constraints.  We were able to execute this scan using just two separate tunings with three spectral windows each (S. Schouws et al.\ 2021, \textit{in prep}).  The full bandwidth for each scan was therefore 10.75 GHz in total.  Approximately 40 minutes were required in each tuning to reach a sensitivity of 0.25 mJy per 66 km/s, which was chosen to ensure detection of [CII]$_{158\mu m}$ based on previous studies \citep[e.g.,][]{Smit_2018Natur.553..178S}.  The precipitable water vapor during the executed observations ranged from 1.6 to 3.5 mm.

Additionally, we also utilize the  sensitive, higher spatial resolution band-6 observations we obtained of $z=6.853$ source COS-3018555981 (hereafter abbreviated to COS-3018).  The purpose of those observations was to characterize in detail the kinematics of a $z\sim7$ galaxy with relatively luminous \cii line and to determine if the kinematics were more consistent with simple rotation or a merging system.  COS-3018 was one of two $z\sim6.8$ sources featured in \citet{Smit_2018Natur.553..178S} and found to show prominent differential motion across the detected \cii profile.  Thanks to the depth of the band-6 observations on this source, we can also use them to characterize the IRX in this source.

In interpreting the ALMA results obtained for our $z\sim7$ galaxy targets, it is worthwhile asking whether our selection criteria for our targets might bias our results.  After all, our selection of $z\sim7$ targets required that their  [OIII]$_{5007\text{\AA}}$+H$\beta$ nebular emission lines be sufficiently strong to narrow the redshift likelihood distribution for sources.  As we will discuss in Schouws et al.\ (2021, \textit{in prep}), the median source in our $z\sim7$ selection only had a [OIII]$_{5007\text{\AA}}$+H$\beta$ EW of $\sim$500\AA.  Since this is consistent with the $759_{-113}^{+112}$ \AA$\,$rest-frame EW derived by \citet{Endsley_2021} for [OIII]$_{5007\text{\AA}}$+H$\beta$ for the $z\sim7$ population, we can conclude that our $z\sim7$ results appear to be representative.

\subsection{ALMA Continuum Observations of Bright $z\sim8$ Galaxies}

We additionally targeted the eight most luminous Lyman Break Galaxies at $z\sim8$ in UltraVISTA to obtain sensitive continuum constraints on their far-IR emission and obscured star formation. The targeted sources were identified as part of a search for bright Lyman-break galaxies at $z\sim8$ \citep{Stefanon2017,Stefanon2019}, where sources were required to have $Y - (J+H)/2$ colors redder than 0.75 mag, to show a blue $UV$-continuum slope redward of the Balmer break, and a low-probability of being a low-redshift galaxy based on SED fits.

The eight sources were targeted as part of two ALMA programs \#2017.1.01217.S (PI Stefanon) and \#2018.1.00236.S (PI Stefanon) and observed for $\sim$35 minutes each.  To optimize the probability for detection for \cii, we used band-6 to obtain continuum observations of those sources whose best-fit photometric redshifts were $z<8$ and band-5 to obtain continuum observations for those whose best-fit redshifts were $z>8$.  As a result of the relatively large photometric redshift uncertainties on the \citet{Stefanon2019} $z\sim8$ sources and the cost of pursuing spectral scans, we elected to use a single spectral setup to maximize the continuum sensitivity.  We also obtained 58 minutes of follow-up observations (\#2018.A.00022.S: PI Schouws) on one source from
ALMA program \#2017.1.01217, UVISTA-Y-004, in an effort to confirm a tentative 2.5$\sigma$ dust-continuum detection in that source.

\begin{figure}[t]
\epsscale{1.17}
\plotone{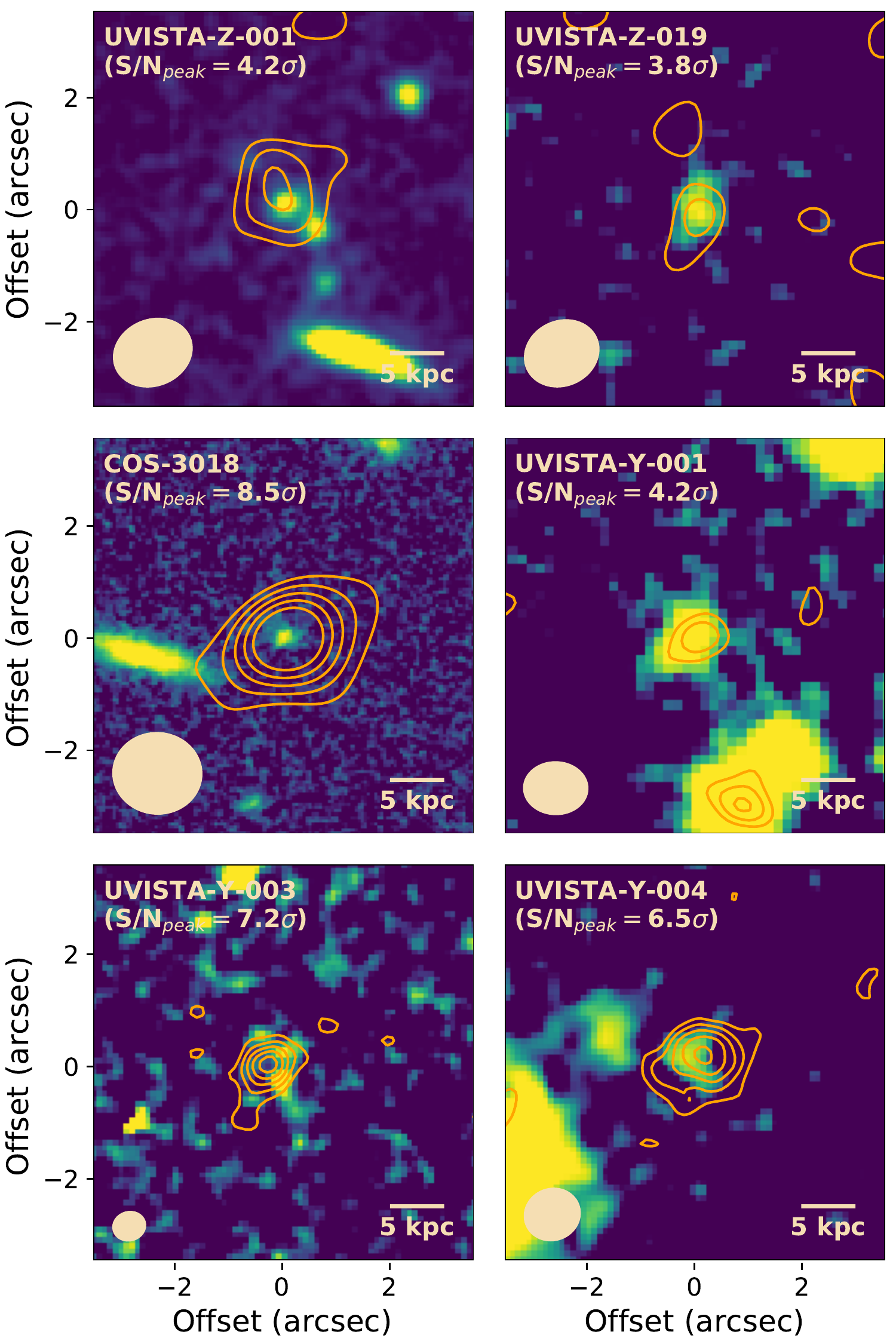}
\caption{An overlay of band-6 dust-continuum observations (1.2$\mu$m) with ALMA on the available near-IR  imaging observations for 6 $z\sim7$-8 sources detected in the dust continuum. For UVISTA-Z-001, UVISTA-Y-003 and COS-3018, near-IR imaging observations are from 1-orbit HST F140W, F160W, and F160W exposures, respectively, and for the remaining sources, from stacked J+H+K$_s$ imaging from UltraVISTA.  Contours are drawn at (2, 3, 4, 5, 6)$\times\sigma$ where $\sigma\approx15\mu$Jy beam$^{-1}$. The ALMA contours presented here are from the natural weighted imaging except for COS-3018 where a 1.5" taper is used (see Section \ref{sec:data_reduction}). The synthesized beam for the dust-continuum observations is indicated in the lower-left corner of each panel. The significance of the dust detections for each source is stated below the sourcename (peak-S/N).  In Figure~\ref{fig:nondetections}, we show the continuum results for sources lacking a clear detection.\label{fig:postage_schouws}}
\end{figure}

\begin{deluxetable*}{lcccccccc}
\tablecaption{Physical and Measured Characteristics of the Bright $z\sim7$-8 Galaxies Targeted with ALMA Observations \label{tab:results}}
\tablewidth{0pt}
\tablehead{\colhead{Sourcename} & \colhead{Redshift\tablenotemark{$\dagger$}} & \colhead{$m_{UV}$}     & \colhead{$\beta_{UV}^{\ddagger}$} & \colhead{\specialcell[c]{$L_{UV}$\\($10^{11} L_{\odot}$)}} & \colhead{\specialcell[c]{log($M_{*}$)$^{\ddagger}$\\($M_{\odot}$)}} & \colhead{\specialcell[c]{$S_{\nu}$$^{b}$\\($\mu$Jy)}} & \colhead{\specialcell[c]{$L_{IR}$$^d$\\($10^{11} L_{\odot}$)}} }
\startdata
   UVISTA-Z-001 &                    7.0611$^a$ &       23.9 $\pm 0.1$ & $-$1.88$^{+0.22}_{-0.07}$ &       2.9$^{+0.1}_{-0.1}$ &    9.58$^{+0.09}_{-0.35}$ &         104$^{+43}_{-43}$ &       5.0$^{+2.1}_{-2.1}$ \\
   UVISTA-Z-007 &                    6.7498$^a$ &       24.5 $\pm 0.1$ & $-$2.02$^{+0.22}_{-0.22}$ &       1.5$^{+0.2}_{-0.2}$ &    9.57$^{+0.35}_{-0.44}$ &                  $<$ 52.2 &                   $<$ 2.2 \\
   UVISTA-Z-009 &    6.90$^{+0.10}_{-0.06}$ &       24.6 $\pm 0.1$ & $-$1.88$^{+0.21}_{-0.51}$ &       1.6$^{+0.2}_{-0.2}$ &    9.40$^{+0.32}_{-0.29}$ &                  $<$ 57.0 &                   $<$ 2.4 \\
   UVISTA-Z-010 &    7.07$^{+0.09}_{-0.07}$ &       24.6 $\pm 0.1$ & $-$2.39$^{+0.35}_{-0.28}$ &       1.1$^{+0.2}_{-0.2}$ &    8.88$^{+0.28}_{-0.09}$ &                  $<$ 44.1 &                   $<$ 2.1 \\
   UVISTA-Z-013 &    7.02$^{+0.06}_{-0.02}$ &       24.9 $\pm 0.2$ & $-$2.24$^{+0.36}_{-0.15}$ &       1.4$^{+0.4}_{-0.3}$ &   10.72$^{+0.03}_{-0.10}$ &                  $<$ 45.0 &                   $<$ 2.2 \\
   UVISTA-Z-019 &                    6.7544$^a$ &       25.1 $\pm 0.2$ & $-$1.51$^{+0.29}_{-0.15}$ &       1.0$^{+0.1}_{-0.1}$ &    9.51$^{+0.19}_{-0.18}$ &          66$^{+23}_{-23}$ &       2.7$^{+0.9}_{-0.9}$ \\
 COS-3018 &                    6.8540$^a$ &       24.9 $\pm 0.1$ & $-$1.22$^{+0.51}_{-0.51}$ &       1.3$^{+0.1}_{-0.1}$ &    9.14$^{+0.18}_{-0.06}$ &          65$^{+13}_{-13}$ &       3.0$^{+0.6}_{-0.6}$ \\
   \hline UVISTA-Y-001 &    8.53$^{+0.53}_{-0.62}$ &       24.8 $\pm 0.1$ & $-$1.37$^{+0.36}_{-0.44}$ &       2.1$^{+0.3}_{-0.3}$ &      10.0$^{+0.9}_{-0.4}$ &          73$^{+20}_{-20}$ &       3.5$^{+1.0}_{-1.0}$ \\
   UVISTA-Y-002 &    8.21$^{+0.50}_{-0.49}$ &       24.8 $\pm 0.2$ & $-$2.60$^{+0.58}_{-0.51}$ &       1.9$^{+0.4}_{-0.3}$ &       9.0$^{+0.3}_{-1.2}$ &                  $<$ 39.9 &                   $<$ 2.3 \\
   UVISTA-Y-003 &    7.62$^{+0.14}_{-0.28}$ &       25.0 $\pm 0.1$ & $-$1.88$^{+0.80}_{-0.73}$ &       1.3$^{+0.2}_{-0.2}$ &       9.9$^{+0.6}_{-0.3}$ &         241$^{+30}_{-30}$ &      10.7$^{+1.3}_{-1.3}$ \\
   UVISTA-Y-004 &    7.42$^{+0.19}_{-0.20}$ &       24.9 $\pm 0.2$ & $-$1.80$^{+0.29}_{-0.29}$ &       1.5$^{+0.4}_{-0.3}$ &       9.9$^{+0.5}_{-0.2}$ &          65$^{+17}_{-17}$ &       2.9$^{+0.8}_{-0.8}$ \\
   UVISTA-Y-005 &    8.60$^{+0.58}_{-0.65}$ &       24.9 $\pm 0.2$ & $-$1.59$^{+1.24}_{-0.58}$ &       1.8$^{+0.5}_{-0.4}$ &       9.0$^{+0.4}_{-1.1}$ &                  $<$ 38.7 &                   $<$ 2.5 \\
   UVISTA-Y-006 &    8.32$^{+0.66}_{-0.92}$ &       25.3 $\pm 0.3$ & $-$1.70$^{+0.70}_{-0.80}$ &       1.2$^{+0.3}_{-0.3}$ &       9.7$^{+1.1}_{-0.5}$ &                  $<$ 39.6 &                   $<$ 2.5 \\
   UVISTA-Y-007 &    8.47$^{+0.73}_{-0.72}$ &       25.5 $\pm 0.3$ & $-$2.00$^{+0.70}_{-0.50}$ &       1.1$^{+0.4}_{-0.3}$ &       ---$^e$ &                  $<$ 53.7 &                   $<$ 2.5 \\
   UVISTA-Y-009 &    7.69$^{+0.99}_{-0.71}$ &       25.4 $\pm 0.3$ & $-$2.60$^{+0.90}_{-0.60}$ &       1.0$^{+0.4}_{-0.3}$ &       ---$^e$ &                  $<$ 53.6 &                   $<$ 2.5 \\ 
   \hline Stack at z$\sim$7$^{c}$        &  6.95$^{+0.07}_{-0.08}$   &       24.6 $\pm 0.2$ & $-$1.99$^{+0.25}_{-0.25}$ &       1.5$^{+0.2}_{-0.2}$ &       9.6$^{+0.6}_{-0.4}$ &                  33$^{+21}_{-24}$ &              1.4$^{+1.0}_{-1.0}$  \\
   Stack at z$\sim$8$^{c}$         &    7.87$^{+0.20}_{-0.23}$ &       25.1 $\pm 0.2$ & $-$2.13$^{+0.21}_{-0.21}$ &       1.5$^{+0.1}_{-0.1}$ &       9.3$^{+0.5}_{-0.9}$ &                  27$^{+14}_{-12}$ &                   1.3$^{+0.6}_{-0.5}$ \\
   \specialcell[c]{Stack of \\ non-detections}        &    7.72$^{+0.40}_{-0.42}$ &       24.9 $\pm 0.2$ & $-$2.16$^{+0.22}_{-0.22}$ &       1.4$^{+0.3}_{-0.3}$ &       9.2$^{+0.3}_{-0.7}$ &                  $<$ 23.5 &                   $<$ 1.2 \\
\enddata
\tablecomments{\textsuperscript{$\dagger$}Quoted redshifts are based on the photometric redshift estimates given in Schouws et al.\ (2021, \textit{in prep}) or \citet{Stefanon2019}. 
\textsuperscript{$\ddagger$}$UV$-continuum slopes $\beta_{UV}$ and stellar masses are from \citet{Stefanon2019} or use the methodology described in \citet{Stefanon2019} for these estimates, assuming a 0.2 $Z_{\odot}$ metallicity a constant star formation prescription, and a \citet{Calzetti_2000} dust law. \textsuperscript{a}Spectroscopic redshifts are from the \cii line (Schouws et al. 2021, \textit{in prep} and \citealt{Smit_2018Natur.553..178S} for COS-3018).
\textsuperscript{b}Integrated flux derived with UVMULTIFIT (not corrected for CMB effects: see \S \ref{sec:individual_results}).  Upper limits on non detections are 3$\sigma$.
\textsuperscript{c}Stack fluxes and errors are determined using a bootstrapping analysis (see section \ref{sec:stacking}).
\textsuperscript{d}Total infrared luminosity integrated from 8-1000$\mu$m assuming a modified black body SED with a dust temperature of 50K and a dust emissivity index $\beta_{dust}$=1.6 after correcting for CMB effects (see \S \ref{sec:individual_results}).
\textsuperscript{e}Due to the challenges in measuring the flux of these sources in the Spitzer/IRAC data due to confusion from bright neighbors \citep{Stefanon2019}, no stellar masses are estimated for these sources.}
\vspace{-0.6cm}
\end{deluxetable*}

\subsection{ALMA Data Reduction} \label{sec:data_reduction}

The ALMA data were reduced and calibrated using \textsc{Casa} version 5.4.1 following the standard ALMA pipeline procedures. The imaging was performed using the \textsc{tclean} task in \textsc{Casa}, applying natural weighting to maximize signal to noise. Any channels containing line emission from the \cii line were carefully excluded from the continuum imaging. All significant continuum sources in the field were masked and cleaned conservatively to 3$\sigma$. A range of \textit{uv}-tapers from 1.0 to 2.0 arcseconds were applied, but it was found that the images without any \textit{uv}-tapering had the highest SNR.  Therefore we use the setup without any \textit{uv}-tapering for the remainder of this paper unless specified otherwise. We reduce the observations for COS-3018 in a similar way except applying a 1.5 arcsecond taper to the visibility data; the full resolution results will be presented in Smit et al. (2021, \textit{in prep}).

To examine in more detail the relative morphologies of galaxies in terms of their far-IR dust-continuum emission and in the rest-frame $UV$, we also re-imaged the sources using  Briggs-weighting with robustness parameter 0.3, where a robustness parameter of $-$2 corresponds to uniform weighting (which maximizes the spatial resolution of the continuum signal) and 2 to natural weighting (which maximizes the S/N).  Our choice of 0.3 for the robustness parameter was made as a compromise between resolution and S/N.  In particular, for the most luminous source UVISTA-Y-003 we used Briggs weighting to decrease the size of our beam from 0.60"$\times$0.52" beam to a 0.45"$\times$0.37". This is useful to constrain the physical alignment between the UV and FIR, which will be further discussed in Section \ref{sec:alignment}.

\subsection{Near-IR Imaging Observations} 

In interpreting the dust-continuum observations for our $z\sim7$-8 sources, it is helpful to compare against the sensitive imaging observations that are available for these targets in the near-IR.  All 14 of our targets have available sensitive $YJHK_s$ imaging observations from the UltraVISTA program, albeit with limited (FWHM$_{PSF}$ $\sim$0.8$''$) spatial resolution.  For a few sources, higher spatial resolution imaging observations exist from HST, in particular for UVISTA-Z-001 \citep[1-orbit in F140W:][]{bowler2017_10.1093/mnras/stw3296}, UVISTA-Y-003 \citep[1-orbit in F160W, DASH:][]{Momcheva_2016} and COS-3018  \citep[1.3-orbits in F160W, CANDELS:][]{Grogin_2011}.

To ensure good registration of the available near-IR observations with our new ALMA-continuum observations, we took advantage of the high astrometric accuracy of the GAIA DR2 catalog \citep{gaia2018A&A...616A...1G} to make minor ($<$0.1$"$) adjustments to the astrometry of the near-infrared imaging observations available for the COSMOS field from UltraVISTA or HST (if available).  As a further check on the astrometric accuracy of the near-IR imaging data, we compared the position of eight lower-redshift dust-continuum detections in our ALMA data with their position in the near-IR imaging data and found good agreement ($<$0.2$"$).

\begin{figure}[t]
\epsscale{1.20}
\plotone{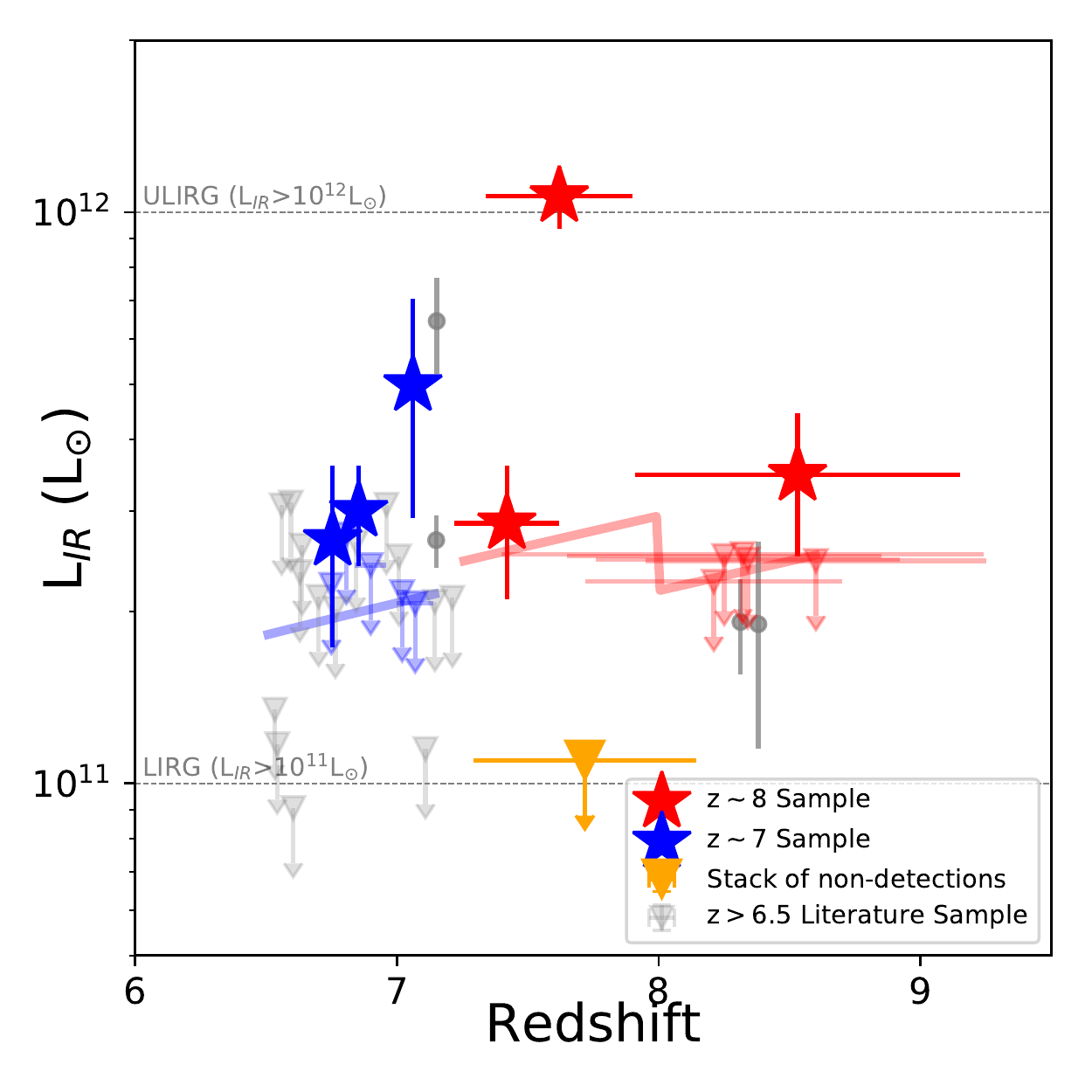}
\caption{IR luminosities and redshifts for $z\sim7$ and $z\sim8$ galaxies detected in ALMA continuum observations analyzed here (\textit{blue and red stars}, respectively).  The redshifts at which sources are shown are either from detected \cii lines (Schouws et al.\ 2021, \textit{in prep}) or from the computed photometric redshifts.  The blue and red plotted downward pointing triangles correspond to $3\sigma$ upper limits on the IR luminosities for the four $z\sim7$ and 5 $z\sim8$ sources undetected in our ALMA observations.  Also shown are previous dust-continuum detections from the literature at $z>6$ \citep[\textit{solid grey circles}:][]{Laporte_2017,Bowler_2018_10.1093/mnras/sty2368,Tamura_2019,Hashimoto_2019_10.1093/pasj/psz049}.  The thick blue and red lines shows the $3\sigma$-limiting IR luminosities we probe given the integration times for our ALMA observations.\label{fig:dustz}}
\end{figure}

\section{Results} \label{sec:results}

\subsection{Individual Detections} \label{sec:individual_results}

The new ALMA observations we have obtained on a significant sample of 15 bright $z\sim7$-8 galaxies provide us with essential new information on the obscured star formation and dust-continuum emission in these sources.

Using naturally-weighted reductions of the continuum observations, we performed a systematic search for 3$\sigma$ peaks within 1$"$ of the rest-$UV$ positions of our 15 targeted sources.  We find a significant dust continuum detection for six of the sources, as is illustrated in Figure \ref{fig:postage_schouws}, which shows the 2, 3, 4, 5, 6$\sigma$ dust-continuum contours overlaid on the rest-$UV$ emission.  The continuum detections show  $<$0.5$"$ spatial offsets from the UV positions. We note that allowing for large spatial offsets (e.g., to 3$"$) does not increase the number of detected sources. 

\begin{figure}[t]
\epsscale{1.19}
\plotone{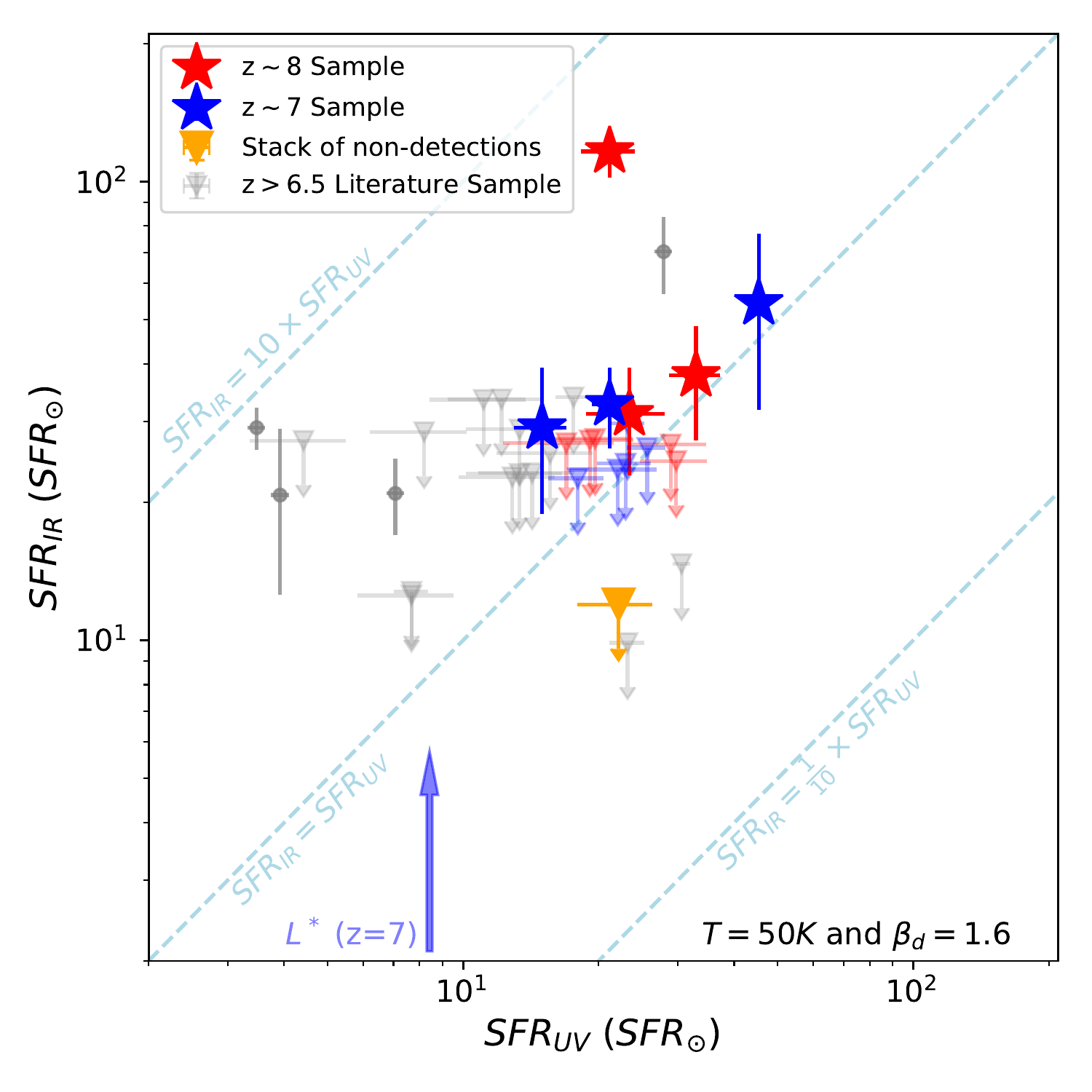}
\caption{A comparison of the star formation rates we infer from new IR-continuum observations we have obtained for bright $z\sim7$-8 galaxies with the star formation rates observed in the rest-$UV$ (\textit{blue and red colored stars}, respectively).  Previous results for $z>6.5$ galaxies are shown with the gray symbols and upper limits  \citep[][]{Ota_2014,Maiolino_2015_10.1093/mnras/stv1194, Watson_2015_Natur.519..327W, Pentericci_2016, Knudsen_2017_10.1093/mnras/stw3066, Laporte_2017,Bowler_2018_10.1093/mnras/sty2368,Hashimoto_2019_10.1093/pasj/psz049,Tamura_2019, Matthee_2017, Matthee_2019}.  Upper limits are $3\sigma$.  For reference, the typical UV star formation rate of L$^*$ galaxies at $z=7$ \citep{Bouwens_2015} is indicated with the blue arrow.  For the $UV$ luminous sources we analyzed here, the obscured SFRs (SFR$_{IR}$) appear to be comparable to the unobscured SFRs (SFR$_{UV}$). \label{fig:sfrirsfruv}}
\end{figure} 

\begin{figure*}[t]
\epsscale{0.9}
\plotone{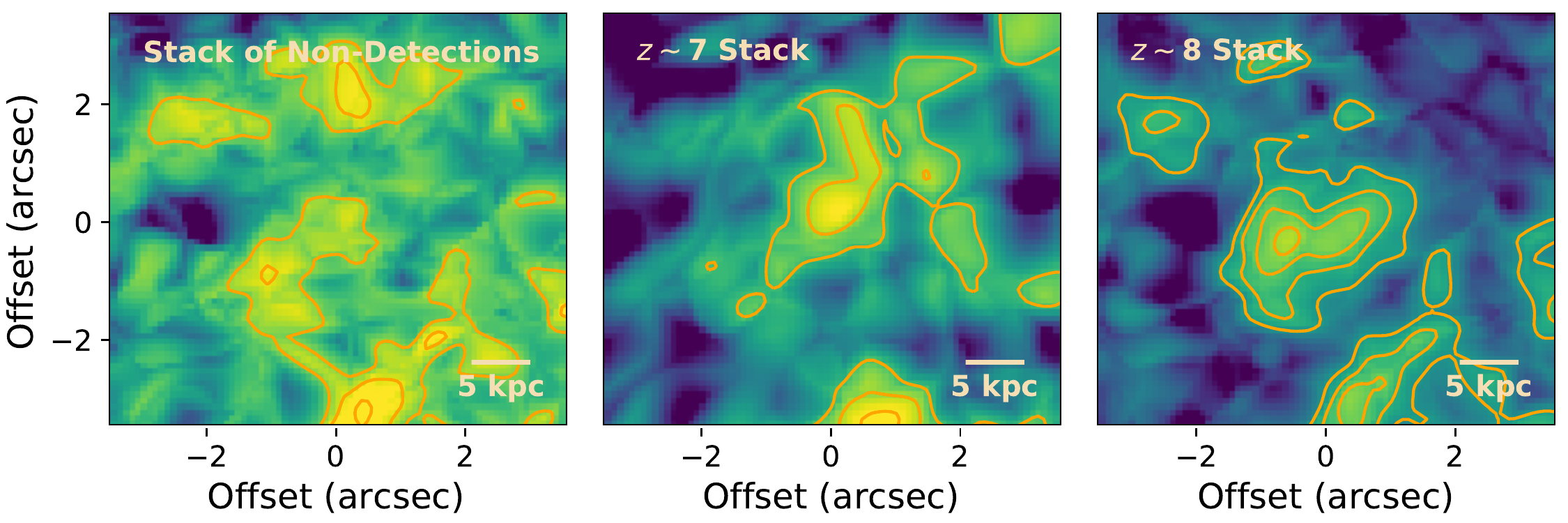}
\caption{\textit{Left:} Median stacking of all non-detections does not result in a significant detection of the dust continuum. Contours are drawn at (1, 2, 3, 4, 5, 6)$\times\sigma$ where $\sigma=6.9\mu$Jy beam$^{-1}$.
\textit{Middle and Right:} Median stacks combining the ALMA observations for all $z$$\sim$7 (middle) and $z$$\sim$8 (right) galaxies from our sample (including detections and non-detections) to study the typical properties of our sample. These stacks are marginally detected at 3.0$\sigma$ and 3.8$\sigma$ respectively (peak-SNR).  See Section \ref{sec:stacking} and \ref{sec:sfrd} for a discussion of these results. \label{fig:stacks}}
\end{figure*}

To measure continuum fluxes for the detected sources, the complex visibilities were fitted using UVMULTIFIT \citep{uvmultifit2014A&A...563A.136M}, applying a 2D Gaussian model with a single component.  By fitting in the (u,v)-plane, the results are independent of the imaging parameters.  The fluxes measured in this way are consistent with measurements in the image plane and measured fluxes range from 75$_{-10}^{+6}$ $\mu$Jy to 302$_{-35}^{+45}$ $\mu$Jy, comparable with previous studies at $z\gtrsim$7 \citep{Watson_2015_Natur.519..327W,Willott_2015,Laporte_2017,Knudsen_2017_10.1093/mnras/stw3066,Bowler_2018_10.1093/mnras/sty2368,Tamura_2019,Hashimoto_2019_10.1093/pasj/psz049}. In the case of non-detections, we provide $3\sigma$ upper limits derived from the noise measured in the natural weighted imaging. For convenience, we summarize the results in Table \ref{tab:results}. The fluxes in this table have not been corrected for effects arising from the Cosmic Microwave Background (CMB). For derived quantities ($L_{IR}$, $M_{dust}$ etc.) we correct the measured fluxes for the reduced contrast due to the higher temperature of the CMB at high redshifts using Eq. (18) of \citet{da_Cunha_2013}.

For comparison with other results in the literature and lower redshift studies, it is convenient to estimate the total infrared luminosities for the sources in our sample.  Following previous studies, we adopt a modified single-temperature black body (greybody) dust spectral energy distribution and integrate the SED between 8-1000$\mu$m to calculate $L_{IR}$. 
For the dust emissivity index we assume $\beta_d$=1.6, which is the best-fitting value for local infrared luminous galaxies \citep{Casey_2014}. 

Dust temperatures in typical star-forming galaxies in the local universe are around $\sim$35 K \citep{Casey_2014,remy-ruyer-2015A&A...582A.121R}, but it has been shown using stacking of Herschel observations that dust temperatures increase with redshift \citep{bethermin_2015A&A...573A.113B,schreiber2015A&A...575A..74S,Schreiber_2018A&A...609A..30S, Faisst_2017, Faisst_2020_10.1093/mnras/staa2545}.  Other observational  \citep[e.g.,][]{J_Bouwens_2016_dust, coppin2014, Knudsen_2017_10.1093/mnras/stw3066, Capak2015_Natur.522..455C,Watson_2015_Natur.519..327W, bouwens_2020} and theoretical \citep{Behrens2018MNRAS.477..552B,Liang2019MNRAS.489.1397L,Sommovigo2021MNRAS.503.4878S} studies also suggest such a trend.  We therefore adopt a dust temperature of 50K for our analysis (this is the dust temperature after any CMB heating).

To put our dust continuum detections in context, we show the IR luminosities we have inferred for the sources in our samples vs. redshift in Figure~\ref{fig:dustz} and then compare these detections with IR luminosities inferred for other dust-detected galaxies at $z>6$.  Interestingly enough, one of our $z>7$ galaxies has an IR luminosity in excess of $10^{12}$ $L_{\odot}$, which places the source in the ULIRG category and is the most IR-luminous galaxy known at $z>7$.  For context, we have also estimated the limiting luminosities to which we would have succeeded in detecting bright $z\sim7$ and $z\sim8$ galaxies, given the integration times for sources in the two samples.  

We estimate $SFR_{UV}$ and $SFR_{IR}$ for sources following \citet{Madau2014ARA&A..52..415M}, with $SFR_{UV} = \kappa_{UV}\cdot L_{UV}$ and $SFR_{IR} = \kappa_{IR}\cdot L_{IR}$ where $\kappa_{UV}=2.5\cdot10^{-10}$$ M_{\odot} yr^{-1} L_{\odot}^{-1}$ and $\kappa_{IR}=1.73\cdot10^{-10}$$ M_{\odot} yr^{-1} L_{\odot}^{-1}$ and we convert from a Salpeter IMF to Chabrier IMF with $SFR_{Cha} = 0.63 \cdot SFR_{Sal}$ \citep{Madau2014ARA&A..52..415M}.
We compare these SFR estimates for our sources to previous detections and upper limits at $z>$6.5 \citep{Ota_2014,Maiolino_2015_10.1093/mnras/stv1194,Watson_2015_Natur.519..327W,Pentericci_2016,Inoue_2016, Knudsen_2017_10.1093/mnras/stw3066, Bradac_2017,Laporte_2017, Bowler_2018_10.1093/mnras/sty2368,Carniani_2018,Hashimoto_2018Natur.557..392H,Smit_2018Natur.553..178S,Hashimoto_2019_10.1093/pasj/psz049,Tamura_2019,Matthee_2019} in Figure~\ref{fig:sfrirsfruv}.  

\begin{figure}[t]
\epsscale{1.17}
\plotone{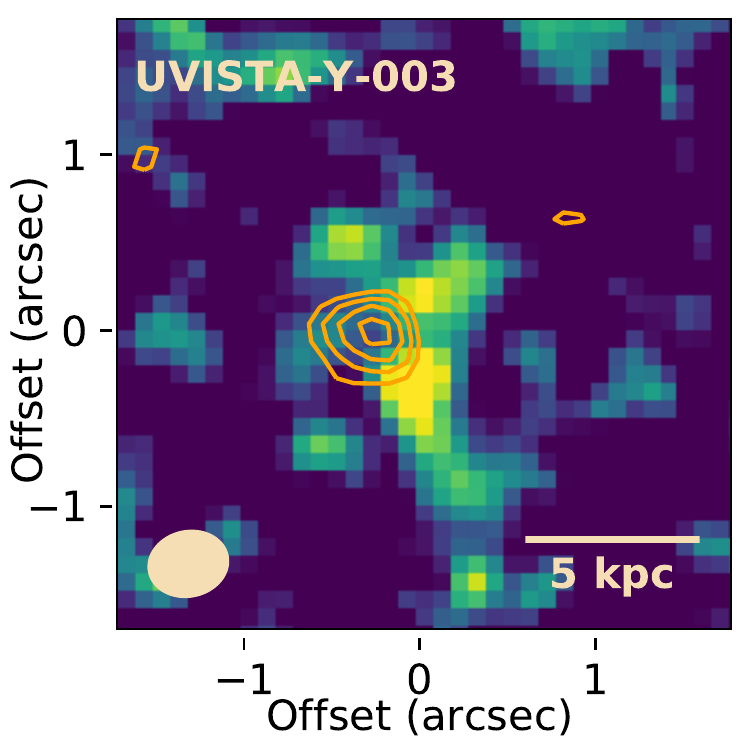}
\caption{Higher spatial resolution image of the dust-continuum emission from the bright $z\sim8$ galaxy UVISTA-Y-003 shown relative to the rest-$UV$ imaging we have of the source with HST in the F160W band (1.6$\mu$m; $\sim$1900\AA$\,$  rest-frame) (3.5"$\times$3.5" FOV).  The presented imaging of UVISTA-Y-003 used a Briggs weighting with robustness parameter 0.3 to bring out the high-spatial resolution structure (Beam: 0.45"$\times$0.37").  Contours are drawn at (2.0, 2.5, 3.0, 3.5)$\times\sigma$ where $\sigma=23\mu$Jy.  Interestingly enough, the $UV$-continuum emission from UVISTA-Y-003 appears to be divided into three separate clumps (see  \citealt{Stefanon2019}). All of these rest-$UV$ clumps show a clear spatial offset from the dust-continuum emission.  It therefore seems very likely that there is a high star-formation region in UVISTA-Y-003, which is entirely obscured in the rest-$UV$, much like one commonly observes in ULIRGs at both $z\sim0$ and $z\sim1$-3.  It is unclear whether the multiple rest-$UV$ components are indicative of merger activity or clumpy star forming regions. \label{fig:postage_highres}}
\end{figure}

Interestingly enough, we find that the obscured star formation rates $SFR_{IR}$ for sources are fairly similar to the unobscured star formation rates $SFR_{UV}$, in cases where the dust is detected.  This is consistent with previous results in the literature which include a few prominent detections where $SFR_{IR}>SFR_{UV}$ \citep{Knudsen_2017_10.1093/mnras/stw3066, Laporte_2017,Bowler_2018_10.1093/mnras/sty2368,Hashimoto_2019_10.1093/pasj/psz049,Tamura_2019} and many sources which lack prominent dust detections and $SFR_{IR}<SFR_{UV}$ \citep[][]{Ota_2014,Maiolino_2015_10.1093/mnras/stv1194, Watson_2015_Natur.519..327W, Pentericci_2016, Matthee_2017, Matthee_2019}.

\subsection{Stacking analysis\label{sec:stacking}}

Nine of the bright $z\sim7$-8 galaxies targeted by our ALMA programs are not individually detected in our continuum observations.  To better characterize dust-continuum emission and obscured star formation in these sources, we can combine the continuum observations on these sources to obtain an average constraint through median-stacking.  In addition to stacking those sources which are individually undetected, we repeat the stacking analysis using our entire $z\sim7$ and $z\sim8$ samples separately, to constrain the properties of the median source in each sample.

For the stacking, we center sources according to their apparent position in the available HST or UltraVISTA rest-UV observations and coadd the ALMA continuum observations after tapering the continuum images with a 1-arcsec taper and and weighting the sources equally. We use tapered images to ensure that we capture all of the dust-continuum flux and to decrease the impact of both (1) the varying resolutions of the ALMA beam and (2) offsets between the dust continuum and UV emission on the stack.  This most relevant for the sources that were observed with a smaller beam, i.e., $<1"$.

The results of the stack of the nine non-detections are presented in both Table~\ref{tab:results} and Figure~\ref{fig:stacks} (\textit{left panel}). No significant ($<$2$\sigma$) detection is found, which translates to a 3$\sigma$ upper limit of 1.2$\times10^{11}L_{\odot}$ (assuming T=50K and $\beta_d$=1.6) on the IR luminosities of the undetected sources (the stack is also not detected when stacking using a weighted mean).  In Table~\ref{tab:results} and Figure~\ref{fig:stacks}, we also present our stack results for our $z\sim7$ and $z\sim8$ samples.  A stack of our $z\sim7$ sample shows a $3.0\sigma$ detection and a stack of our $z\sim8$ sample shows a $3.8\sigma$ detection.  The mean IR luminosities we infer for bright $z\sim7$ and $z\sim8$ is 1.4$_{-1.0}^{+1.0}$$\times$$10^{11}$ $L_{\odot}$ and 1.3$_{-0.5}^{+0.6}$$\times$$10^{11}$ $L_{\odot}$, respectively.  These results and those presented in Table~\ref{tab:results} are derived based on a bootstrap resampling procedure, re-stacking sub-sets with replacement of our sources 10.000$\times$ and using the median peak flux measured in the stacked images.  The uncertainties are based on the derived 68$\%$ spread in the stack results.

\subsection{Physical alignment between UV and FIR}\label{sec:alignment}

An important longstanding question regards the possible presence of spatial offsets between the dust-continuum emission and the rest-$UV$ emission, as have frequently been found when considering IR luminous galaxies \citep[][]{Smail_2014,dunlop2016,Aravena_2016,laporte2019}.  Indeed, one would expect significant spatial offsets between rest-$UV$ and $IR$-continuum emission in galaxies with a non-uniform dust covering fraction, especially when the optical depths become non-negligible.

To investigate whether similar spatial offsets are found amongst the bright $z\sim7$-8 galaxies we observe with ALMA, we reimage the sources using Briggs weighting, as described in \S2.3, with a beam of 0.45"$\times$0.37" to examine sources at higher  spatial resolution.  Figure \ref{fig:postage_highres} illustrates how our most prominent dust-continuum detection UVISTA-Y-003  compares to the available rest-$UV$ imaging from HST.

For the majority of our sources, the UV and FIR continuum seem well aligned, with the exception of UVISTA-Y-003, which is offset by $\sim$ 0.3 arcseconds ($\sim$2 kpc: see Figure \ref{fig:postage_highres}).  Furthermore, for this source the rest-$UV$ emission seems to be divided into three separate clumps, each of which are offset from the dust emission. The $UV$ clumps are discussed in more detail in \citet{Stefanon2019}, who performed simulations to assess the robustness of the clumps and also confirmed high-$z$ solutions for the clumps based on deblended photometry. Because of these offsets between the $UV$ clumps and the dust continuum emission, it seems likely that there is a high star formation region in UVISTA-Y-003, which is almost entirely obscured in the rest-$UV$, similar to what is observed in ULIRGs at $z\sim0$ and $z\sim1$-3.  It is unclear whether the three rest-$UV$ components indicate the presence of merging activity or not.  It is worth emphasizing the value of the high resolution space-based imaging for UVISTA-Y-003 in revealing both the clumpy nature of this source and the offset between the UV and dust continuum emission.

\subsection{Infrared Excess versus UV continuum Slope} \label{sec:irx}

The measured infrared excess in galaxies ($IRX=L_{IR}/L_{UV}$) is well established to show a correlation with the $UV$-continuum slope $\beta_{UV}$ (where $f_{\lambda}\propto\lambda^{\beta_{UV}}$).  \citet{Meurer_1999} showed that the correlation between the infrared excess and $\beta_{UV}$ could be expressed as follows:

\begin{equation}
    A_{1600} = \frac{dA_{UV}}{d\beta_{UV}}\cdot (\beta_{UV}-\beta_{UV,intr})
\end{equation}

\begin{equation}
    IRX = 1.7 \cdot (10^{0.4\cdot A_{1600}}-1) 
\end{equation}
where $dA_{UV}/d\beta_{UV}$ expresses the steepness of the attenuation law and $\beta_{UV,intr}$ represents the unattenuated $UV$-continuum slope of star-forming galaxies. The IRX-$\beta$ relation has been used as a constraint on the dust attenuation curve of galaxies. In the local universe, starburst galaxies have been shown \citep{Meurer_1999} to be well described by a Calzetti attenuation law \citep{Calzetti_2000}, with $dA_{UV}/d\beta_{UV} = 1.99$ with $\beta_{UV,intr}=-2.23$.  However, subsequent studies have found that some galaxies at higher redshifts  do not follow Calzetti dust law and  tend to favour a steeper attenuation curve, similar to the Small Magellanic Cloud (SMC) for which $dA_{UV}/d\beta_{UV}=1.1$ and $\beta_{UV,intr}$=$-$2.23 \citep{Reddy_2006,Reddy_2018,Siana_2008,Capak2015_Natur.522..455C,J_Bouwens_2016_dust,Barisic_2017,Fudamoto2020arXiv200410760F}.

\begin{figure}[t]
\epsscale{1.2}
\plotone{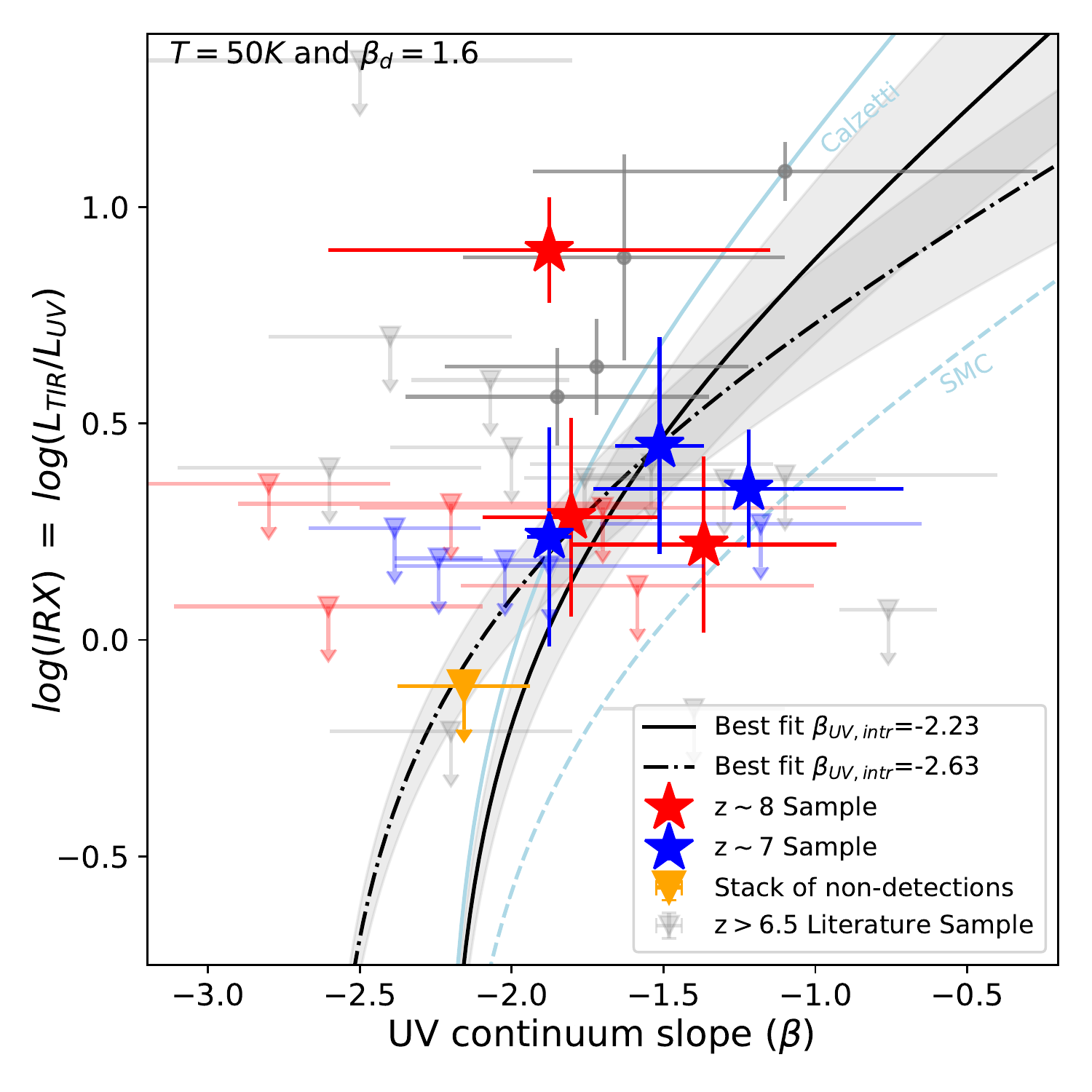}
\caption{The inferred infrared excess of $z>6.5$ galaxies vs. their $UV$-continuum slope $\beta_{UV}$.
The three blue and red stars show the infrared excesses inferred for the $z\sim7$ and $z\sim8$ galaxies
that are detected in the dust continuum.  The blue and red downward-pointing triangles show the $3\sigma$ upper limits on the inferred infrared excess in the eight $z\sim7$ and $z\sim8$ galaxies, respectively, lacking dust-continuum detections.  The orange downward pointing triangle shows the $3\sigma$ upper limit we infer on the infrared excess stacking the individually undetected $z\sim7$-8 galaxies considered here.  The grey circles and downward pointing triangles correspond to dust-continuum detections from the literature \citep[e.g.,][]{Watson_2015_Natur.519..327W,Knudsen_2017_10.1093/mnras/stw3066,Laporte_2017,Bowler_2018_10.1093/mnras/sty2368,Tamura_2019,Hashimoto_2019_10.1093/pasj/psz049}.  For reference, the Calzetti and SMC IRX-$\beta$ relations are shown with the cyan solid and dashed line, respectively (assuming an intrinsic $UV$-continuum slope $\beta_{UV}$ of $-2.23$). \label{fig:IRX-beta}}
\end{figure}

In Figure \ref{fig:IRX-beta}, we present our measurements for the infrared excess (IRX $\equiv$ $L_{TIR}/L_{UV}$) versus the UV continuum slope ($\beta_{UV}$) for $z>6.5$ sources with new measurements from our ALMA programs.  In determining the best-fit value for $dA_{UV}/d\beta$, we consider two different values for $\beta_{UV,intr}$: (1) $\beta_{UV,intr} = -2.23$, as used in modeling the IRX-$\beta$ relation at $z\sim0$ \citep{Meurer_1999} and (2)
$\beta_{UV,intr}=-2.63$, as expected for a younger, lower-metallicity stellar population \citep{Reddy_2018} seen at higher redshift.  For these two values of  $\beta_{UV,intr}$ we determine the best-fit value for $dA_{UV}/d\beta$ for a dust temperature of 50 K and dust emissivity index $\beta_d = 1.6$.  For these fits we use the detections and non-detections described in this paper and a compilation from the literature  \citep[M$_{UV}<$$-$21.5;][]{Maiolino_2015_10.1093/mnras/stv1194,Pentericci_2016,Inoue_2016,Matthee_2017,Matthee_2019,Hashimoto_2019_10.1093/pasj/psz049} and determine the best fit value and errors using a total least squares method which takes account of errors in both dependent and independent parameters.

\begin{figure}[t]
\epsscale{1.20}
\plotone{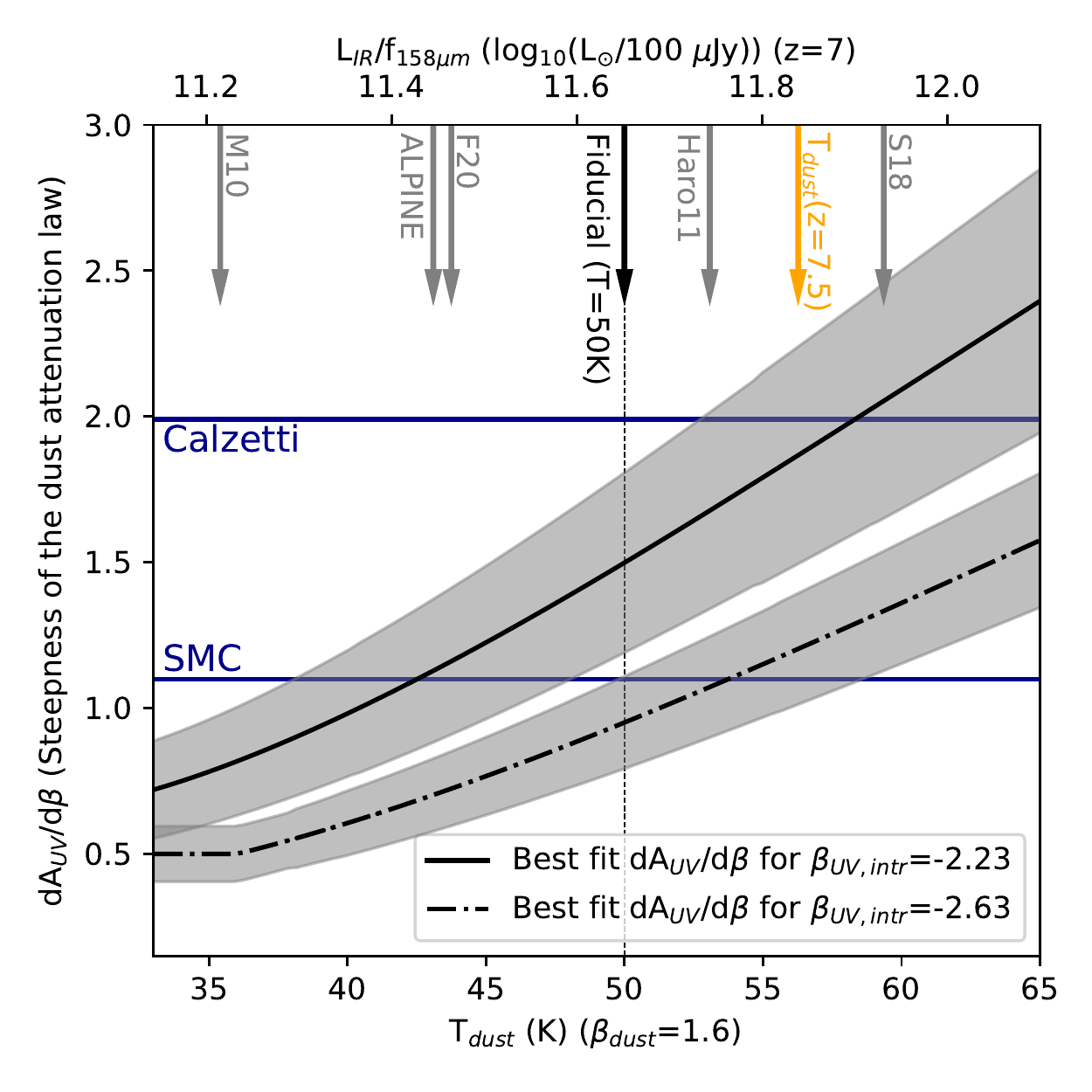}
\vspace{-0.5cm}
\caption{Dependence of the slope parameter of the IRX-$\beta$ relation (dA$_{UV}$/d$\beta_{UV}$) on dust temperature for two different values for the intrinsic $UV$-continuum slope $\beta_{UV,intr}$ (\textit{solid and dash-dotted black lines}).  The grey regions show the slopes dA$_{UV}$/d$\beta_{UV}$ preferred at 68\% confidence.  The fit uses all detections and upper limits presented in this paper and the literature ($z>6.5$ and M$_{UV}<$$-$21.5: $\gtrsim$1.7$L_{UV}^*$).   The blue lines indicate a reference to slope of the SMC and Calzetti relations, while the orange arrow shows the expected dust temperature of LBG galaxies at $z\sim7$-8 extrapolating dust temperature results in the literature \citep{bouwens_2020}. The grey arrows indicate the $L_{IR}$/$f_{158\mu m}$ conversion factors for a variety of high-redshift dust SED templates from the literature and at the corresponding temperature of the modified black body that has the same conversion factor. Here M10 denotes the FIR-SED model from \citet{michalowski2010A&A...514A..67M}, ALPINE the model from \citet{bethermin2020A&A...643A...2B}, F20 from \citet{Faisst_2020_10.1093/mnras/staa2545}, Haro11 from \citet{De_Rossi_2018}, and S18 from \citet{Schreiber_2018A&A...609A..30S}.  The black arrow and vertical dotted line indicate the fiducial 50 K temperature we adopt for our analysis.  For $\beta_{UV,intr}=-2.23$, we find a much better fit of the observational results for an attenuation curve between Calzetti and SMC, and for a $\beta_{UV,intr}=-2.63$, we find a much better fit to a SMC-like attenuation curve.
\label{fig:slope dependence}}
\end{figure}

In Figure \ref{fig:slope dependence}, we generalize the fit results presented in Figure~\ref{fig:IRX-beta} for $dA_{UV}/d\beta$ to allow for an arbitrary dust temperature, given the current uncertainties in establishing the typical dust temperatures for $z>6$ galaxies.  The slopes ($dA_{UV}/d\beta$) of the SMC and Calzetti attenuation curves are shown with horizontal blue lines.  On Figure \ref{fig:slope dependence}, we indicate with a black arrow (and dotted vertical line) the fiducial dust temperature we assume 50 K.  We also indicate with an orange arrow the $T_{dust}$ determination of 54.3$\pm$1.6 K that \citet{bouwens_2020} derived by extrapolating the results from \citet{bethermin_2015A&A...573A.113B},  \citet{Strandet_2016}, \citet{Knudsen_2017_10.1093/mnras/stw3066},
\citet{Schreiber_2018A&A...609A..30S}, \citet{Hashimoto_2019_10.1093/pasj/psz049}, \citet{Bakx_2020}, \citet{bethermin2020A&A...643A...2B}, \citet{Harikane_2020}, and \citet{Faisst_2020_10.1093/mnras/staa2545}.  Also shown on Figure \ref{fig:slope dependence} are where other high-redshift SED templates would lie \citep{bethermin_2015A&A...573A.113B,Schreiber_2018A&A...609A..30S,Faisst_2020_10.1093/mnras/staa2545,michalowski2010A&A...514A..67M,De_Rossi_2018}, including Haro11, which has been argued by \citet{De_Rossi_2018} to be a good low redshift analogue for LBGs and has a relatively high dust temperature, and the ALPINE SED template  \citep{bethermin2020A&A...643A...2B}, which is based on a stacking analysis of the ALPINE target and analogue galaxies fitted to models from \citet{bethermin2017A&A...607A..89B}. The \citet{michalowski2010A&A...514A..67M} template is based on the average SED from a sample of submillimeter galaxies (SMGs). Finally we also indicate the approximate conversion factor and dust temperatures found by \citet{Faisst_2020_10.1093/mnras/staa2545} analyzing a small sample of $z\sim 5.5$ galaxies using the \citet{Casey_2018} models.

For our fiducial $z\sim7$ dust temperature of 50 K and assuming $\beta_{UV,intr}=-2.63$, the current results appear to be more consistent ($3\sigma$ signifiance) with a SMC-like attenuation law than a Calzetti-like attenuation law, while for a $\beta_{UV,intr}=-2.23$, current results lie somewhere between a Calzetti and SMC-like attenuation law.  The present results are similar to earlier $z>6.5$ results by \citet{Watson_2015_Natur.519..327W}, \citet{Laporte_2017}, \citet{Knudsen_2017_10.1093/mnras/stw3066}, \citet{Bowler_2018_10.1093/mnras/sty2368}, and \citet{Tamura_2019} who also found greater consistency with a Calzetti-like attenuation law when assuming $\beta_{UV,intr}=-2.23$.

\begin{figure}[t]
\epsscale{1.18}
\plotone{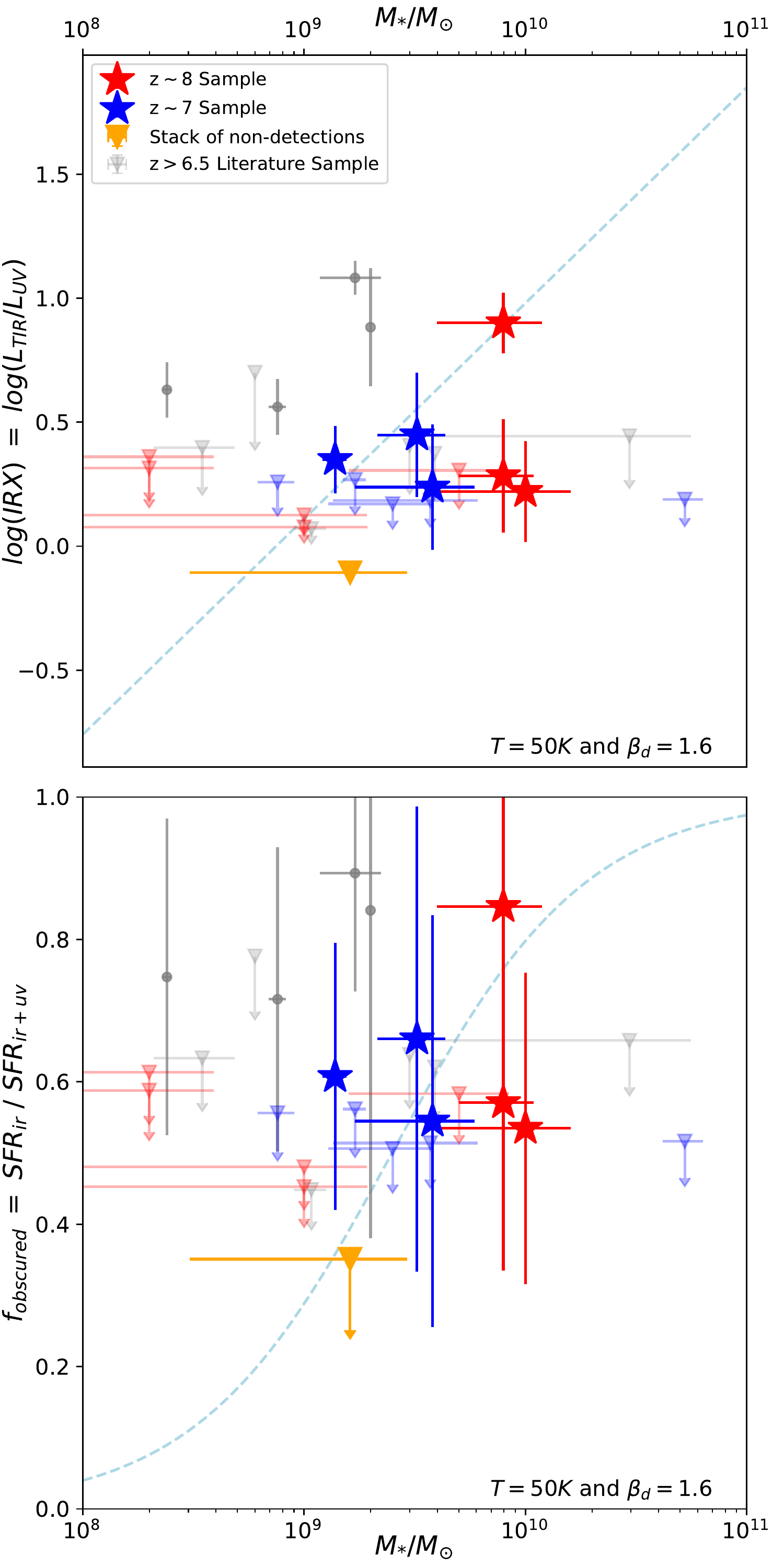}
\caption{(\textit{upper}) IRX-stellar mass relation for a sample of sources at $z>6.5$, including the detections and non-detections presented in this paper and a compilation of sources from the literature \citep{Watson_2015_Natur.519..327W, Knudsen_2017_10.1093/mnras/stw3066, Laporte_2017,Bowler_2018_10.1093/mnras/sty2368,Hashimoto_2019_10.1093/pasj/psz049,Tamura_2019}. The local relation \citep{Whitaker_2017} is shown with the dotted line. \textit{(lower)} Fraction of the star formation that is obscured by dust as a function of stellar mass for our sample and sources from the literature. For a modified black body with a dust temperature of 50K and a dust emissivity of $\beta_{dust}$=1.6 \citep{casey2012_10.1111/j.1365-2966.2012.21455.x}, the $z=0$ relation \citep{Whitaker_2017} is broadly consistent with the detections seen at high redshift. \label{fig:IRX-Mstar_2}}
\end{figure}

\begin{figure}[t]
\epsscale{1.20}
\plotone{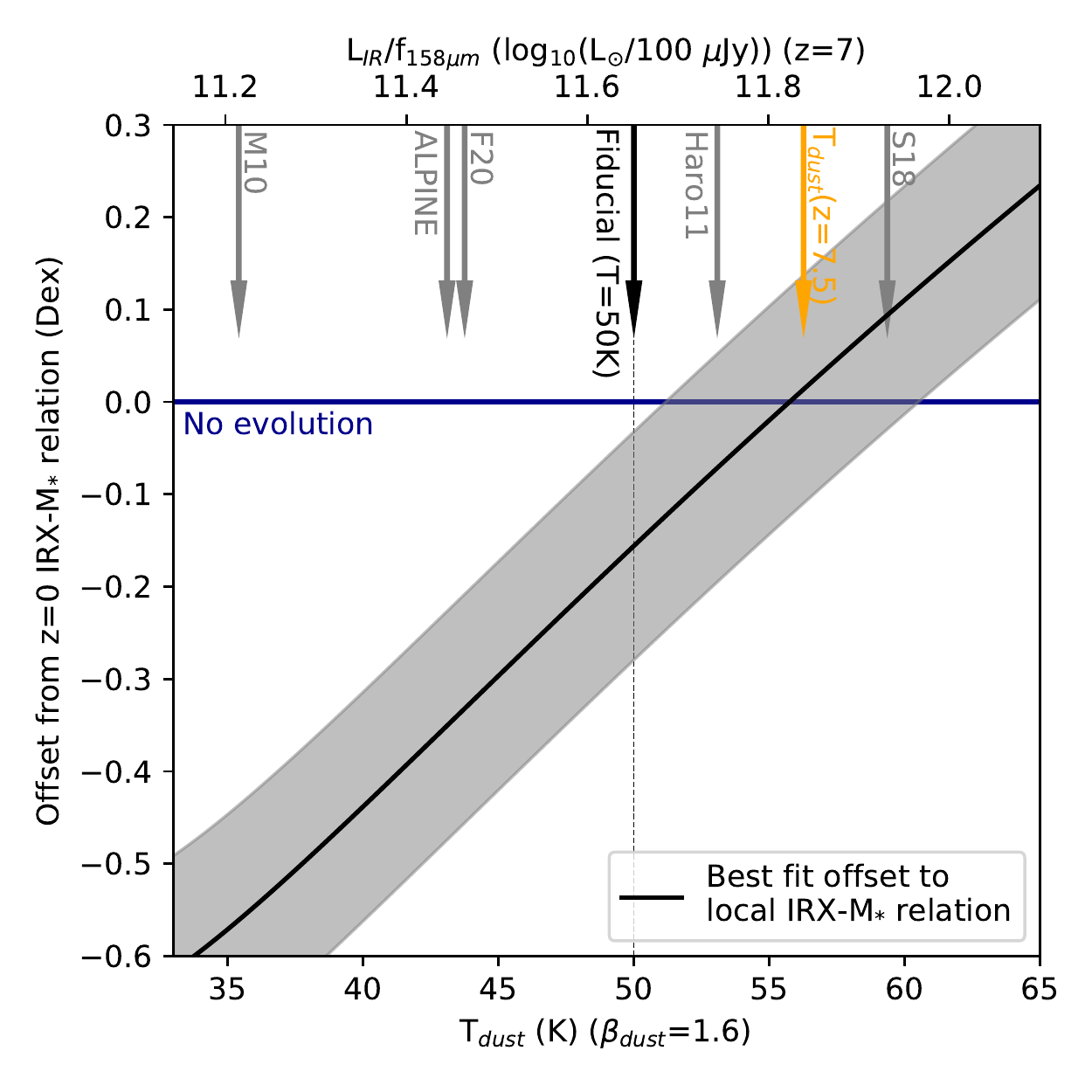}
\caption{Similar to Figure \ref{fig:slope dependence}, this figure shows the dependence on dust temperature of the offset with respect to the local IRX stellar mass relation fitted to the detections and upper limits presented in this paper and the literature ($z>$6.5 and M$_{UV} <$$-$21.5 mag: $\gtrsim$1.7$L_{UV}^*$).  The black line shows the best-fit offset while the grey-shaded area showed the offsets preferred at 68\% confidence.  The blue line is a visual reference corresponding to no evolution.  For fit values below this line, the implied IRX at $z>6.5$ is lower than the local relation at a fixed stellar mass. For the current data, this would be the case for dust temperatures below $\lesssim$54K.  The black, grey, and orange arrows are as in Figure \ref{fig:slope dependence}.  For dust temperatures of $\sim$50-55 K, the present IRX-stellar mass results appear to be broadly consistent with the $z\sim0$ relation \citep{Whitaker_2017}. \label{fig:IRX MSTAR slope dependence}}
\end{figure}

\subsection{IRX-Stellar Mass relation}

The infrared excess is also known to show a strong correlation with the stellar mass of sources, with an approximately linear power-law dependence on stellar mass \citep[e.g.,][]{Bouwens_2016, dunlop2016, mclure2018_10.1093/mnras/sty522}. Interestingly enough, little evolution is observed between $z\sim3$ and $z\sim0$ in the relationship between the infrared excess and the stellar mass \citep{Pannella_2015, J_Bouwens_2016_dust, Whitaker_2017, mclure2018_10.1093/mnras/sty522}.

In Figure~\ref{fig:IRX-Mstar_2} (\textit{upper panel}), we present the infrared excess we derive for our sources vs. their inferred stellar masses.  The blue and red stars correspond to $z\sim7$ and $z\sim8$ galaxies, respectively, detected in the dust-continuum, while the light blue and light red upper limits correspond to sources which are not detected in the dust continuum ($<3\sigma$).  The grey circles show dust-continuum detections from the literature.  Stellar masses for bright $z\sim8$ sources in our selection are as inferred in \citet{Stefanon2019}.  Stellar masses for bright $z\sim7$ sources are inferred using the same SED-fitting assumptions as \citet{Stefanon2019} use at $z\sim8$, i.e., with a 0.2 $Z_{\odot}$ metallicity, constant star formation rate, and \citet{Calzetti_2000} dust curve.  The lower panel in Figure~\ref{fig:IRX-Mstar_2} shows fraction of obscured star formation vs. stellar mass.  Our results seem fairly consistent with the $z\sim0$ relation presented in \citet{Whitaker_2017}.

Similar to Figure~\ref{fig:slope dependence}, we illustrate in Figure~\ref{fig:IRX MSTAR slope dependence} how the normalization in the IRX-stellar mass relation depends on the assumed dust temperature of $z\sim7$-8 galaxies. We quantify the normalization in terms of an offset (in dex) from the low-redshift relation \citep{Whitaker_2017} and fit this offset as a function of dust temperature in a similar way as described in Section \ref{sec:irx}. As in Figure~\ref{fig:slope dependence}, the vertical orange line indicates the expected temperature for $z\sim7$-8 galaxies when extrapolating lower redshift results.  At this temperature, the offset we derive is consistent with no evolution from $z\sim0$.  Not surprisingly, for lower and higher values of the dust temperature, the IRX is lower and higher respectively than the $z\sim0$ relation.

\section{Discussion} \label{sec:discussion}

\subsection{Constraints on Dust Evolution Models}

Our new observations provide constraints on the build-up of dust in early high-redshift galaxies.  In addition to the significant impact dust has on the fraction of $UV$ light seen from star-forming galaxies and that re-emitted in the far-infrared, dust can play a critical role in galaxy evolution \citep{Salim_2020}.  Significantly, the surfaces of dust grains serve as the main production sites of many molecular species and can catalyze the formation of molecular clouds in which star formation (SF) takes place \citep[e.g.,][]{gould1963ApJ...138..393G, Cazaux_2004, hirashita_2002,Yamasawa_2011, chen2018}.  Additionally, dust cooling induces fragmentation inside molecular clouds, which has a significant impact on the stellar initial mass function  \citep[IMF; ][]{Omukai_2000, Omukai_2005,schneider_2006}.

\begin{figure}[t]
\epsscale{1.15}
\plotone{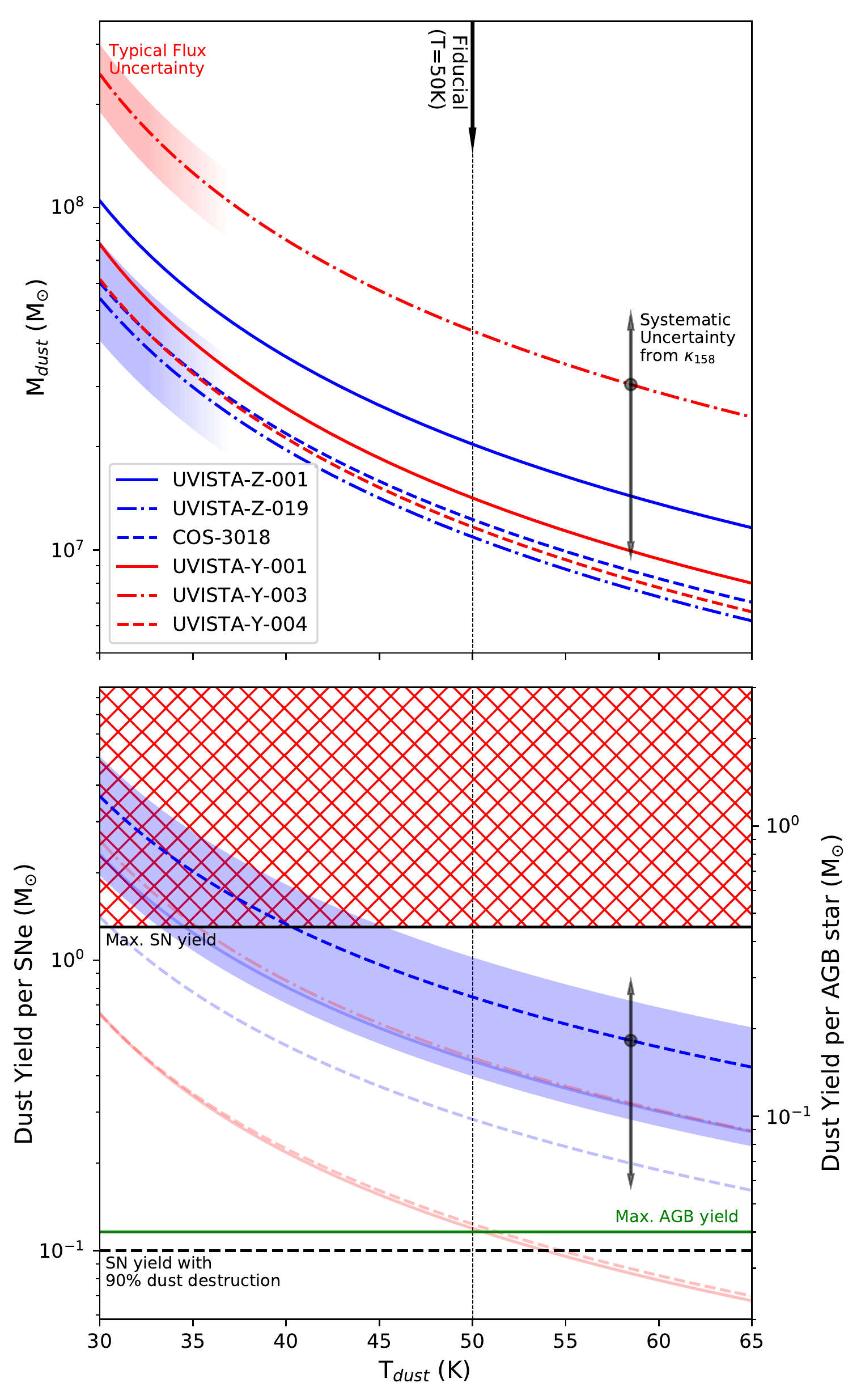}
\vspace{-0.3cm}
\caption{(\textit{upper}) Dust masses as a function of dust temperature. The impact of uncertainties on the flux measurements for the $z\sim$7 and $z\sim$8 samples are indicated with the blue and red shaded regions, respectively. An indicative systematic uncertainty resulting from the choice of dust mass absorption coefficient is shown with the black arrow.  The dotted vertical line indicates the fiducial dust temperature adopted here. (\textit{lower}) Constraint on the dust yield per SNe (\textit{left axis}) or AGB star (\textit{right axis}) as a function of temperature for our sources. We highlight COS-3018 which provides the most demanding constraints on the yields, where the blue shaded region shows the error from the flux measurement and the uncertainty in stellar mass. The black arrow indicates the systematic uncertainty resulting from the choice of dust mass absorption coefficient. The maximum theoretical dust yield from SNe without any dust destruction and after 90\% dust destruction are indicated with the solid and dashed black lines respectively. The maximum yield per AGB star is indicated with the green solid line. Note that if the dust yield constraint for a source is above any of these lines, that particular dust production mechanism alone cannot explain the amount of dust we observe. In particular, for dust temperatures below T$\sim$40K and our assumed value of $\kappa_{158}$, SNe, even without dust destruction, cannot explain the observed dust masses (indicated with the red hatched region).\vspace{-0.7cm} \label{fig:mdust}}
\end{figure}

\begin{figure}[t]
\epsscale{1.20}
\plotone{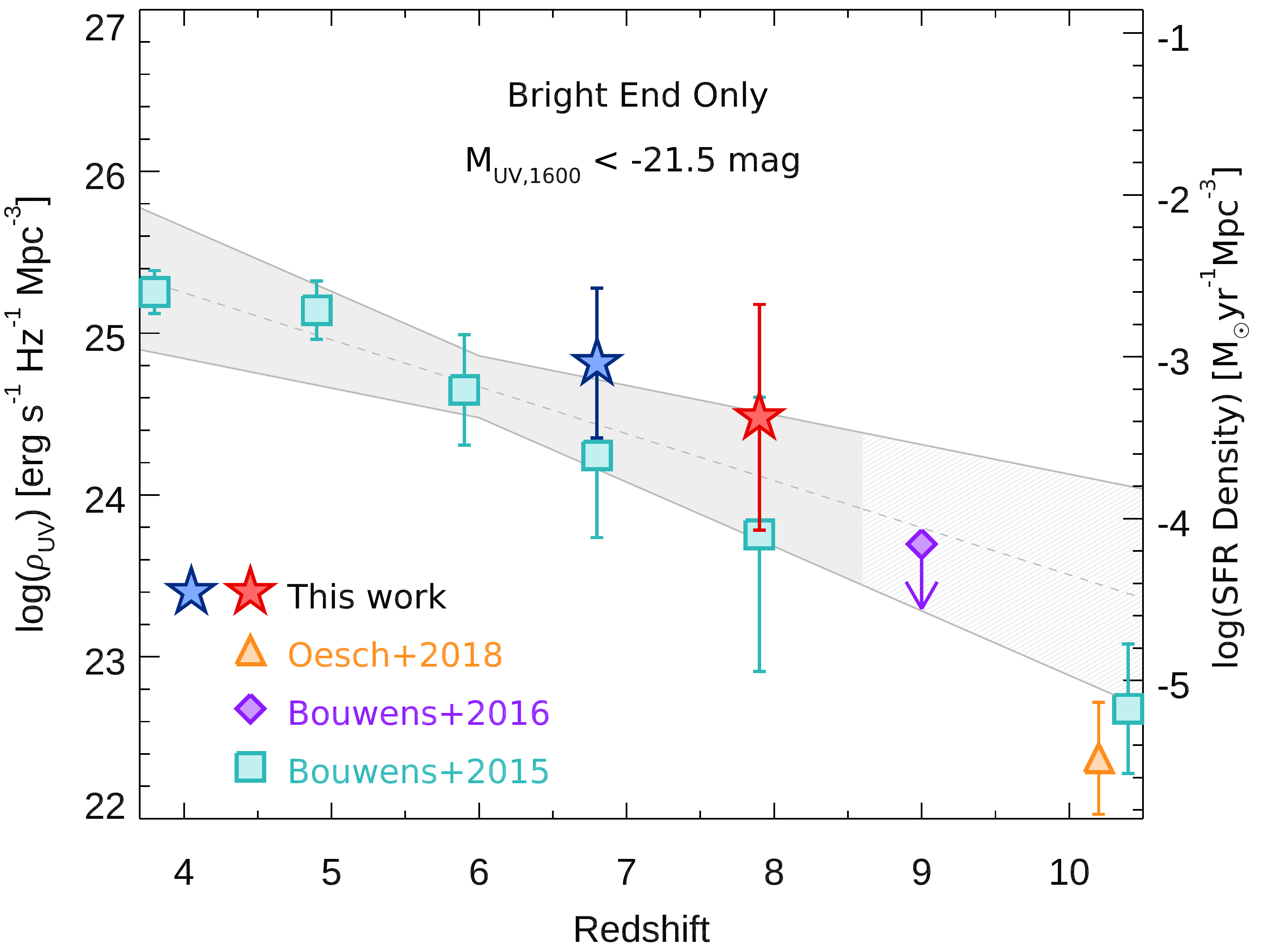}
\caption{The SFR density inferred for especially bright ($M_{UV,AB}<-21.5$ mag: $\gtrsim$1.7$L_{UV} ^*$) galaxies at $z\sim4$-9, based on the rest-$UV$ light (blue circles and squares) alone and including the contribution of dust obscured SF at $z\sim7$ and $z\sim8$ (\textit{red points}).  The shaded grey region indicates the best-fit trend in $UV$ luminosity density (68\% confidence intervals) for $UV$-bright ($<$$-$21.5 mag: $\gtrsim$1.7$L_{UV} ^*$) galaxies vs. redshift \citep{Stefanon2019}.  Our ALMA results suggest that $\sim$25-50\% of the star formation activity in such bright $z\sim7$-8 galaxies is obscured by dust.\label{fig:sfrd}}
\end{figure}

In Figure \ref{fig:mdust}, we show estimates for the dust masses of our sources as a function of temperature, and a strong dependence on the assumed dust temperature can be clearly seen.  We derive our estimates for the dust masses of our sources following the procedure described in \citet{Ota_2014}:
\begin{equation}
   M_{dust} = \frac{S_{\nu_{rest}}D_{L}^{2}}{(1+z)\kappa_{d}(\nu_{rest})B_{\nu_{rest}}(T_{dust})}
\end{equation}
where $S_{\nu}$ is the observed flux density after correction for contrast with the CMB \cite[][equation 18]{da_Cunha_2013},  $D_{L}$ is the luminosity distance, $B_{\nu}(T_{dust})$ is the Planck function at a rest-frame frequency $\nu_{rest}$ and dust temperature $T_{dust}$, and finally $\kappa_{d}(\nu_{rest})$ is the dust mass absorption coefficient, which scales as $\kappa_{d}(\nu_{rest})$ = $\kappa_0 (\nu_{rest}/\nu_{ref})^{\beta_{dust}}$.  Here we assume $\kappa_{0}$ = 8.94 $cm^2$ $g^{-1}$ at 158 $\mu$m which is appropriate for dust ejected from SN after reverse shock destruction \citep{hirashita2014_10.1093/mnras/stu1290}, since it is expected that the majority of dust is produced by SN \citep{michalowski2015A&A...577A..80M,lesniewska2019A&A...624L..13L}. Alternatively, values ranging from $\kappa_{0}$ = 28.4 $cm^2$ $g^{-1}$ at 158 $\mu$m for dust consisting of amorphous carbon to $\kappa_{0}$ = 5.57 $cm^2$ $g^{-1}$ at 158 $\mu$m for dust condensed in SN before reverse shock destruction \citep{hirashita2014_10.1093/mnras/stu1290} could be considered, leading to an additional $\sim$0.7 dex systematic uncertainty. The impact of these systematic uncertainties is shown with the black arrow on Figure \ref{fig:mdust}. Because our data are observed at $\sim$158 $\mu$m rest-frame, which is the same as the calibration wavelength of the dust mass absorption coefficient, our choice of $\beta_{dust}$ introduces minimal uncertainty in the derived dust masses.  While most of our galaxies appear consistent with dust masses of $\sim 10^{7} M_{\odot}$, the large systematic uncertainty limits us to inferring that the dust mass lies somewhere in the range $\sim 10^{7} M_{\odot}$ to $10^{8} M_{\odot}$ for our sources.

In the bottom panel of Figure \ref{fig:mdust} we use these dust masses in combination with the stellar mass from SED fitting to derive the required dust yields per SNe or AGB star following the method described by \citet{michalowski2010A&A...514A..67M}. The stellar mass is translated to the number of SNe or AGB stars that could have contributed to the formation of dust assuming a \citet{Chabrier_2003} IMF and a maximum age of 500 Myr. This number is subsequently divided by the dust mass to obtain the required yields. We compare the required yields for our observed dust masses to the maximum yield from a single AGB star \citep[$\sim$0.004 M$_{\odot}$:][]{morgan2003_10.1046/j.1365-8711.2003.06681.x,ferrarotti_2006A&A...447..553F,ventura_2012_10.1111/j.1365-2966.2012.21403.x,nanni_2014_10.1093/mnras/stt2348,schneider_2014_10.1093/mnras/stu861} and a single supernova without any dust destruction \citep[approximately $\sim$1.3 M$_{\odot}$:][]{todini_10.1046/j.1365-8711.2001.04486.x,Nozawa_2003}. However, it is likely that a large portion of the dust produced by supernovae is destroyed by internal shocks, realeasing only $\lesssim$0.1$M_{\odot}$ into the ISM \citep{bianchi2007_10.1111/j.1365-2966.2007.11829.x,Cherchneff_2010,gall2011A&ARv..19...43G,Laki_evi__2015}.

From this figure it is clear that for our sources dust production from supernovae alone could explain the majority of the observed dust, but only when dust destruction is low (less than $\sim$90\%). Production from AGB stars by itself (green line) is insufficient to explain the observed dust masses for all dust temperatures. For low dust temperatures T$<$40K even dust production from supernovae without any dust destruction would be insufficient and other mechanism like grain growth \citep{michalowski2015A&A...577A..80M, lesniewska2019A&A...624L..13L} would be necessary to explain the observed dust masses.

\subsection{Impact on Measurements of the SFR Density and the IR luminosity function at $z>6$} \label{sec:sfrd}

The ALMA observations from our programs have particular value in determining the quantity of the star formation from $UV$ bright galaxies at $z\sim7$-8 that is obscured by dust.  This is particularly useful for gaining a handle on the likely build-up of mass that occurs in the brightest and most massive galaxies with cosmic time.  

To estimate the approximate correction that we need to make to the SFR density of bright $z\sim7$-8 galaxies to correct for the impact of dust, we compute an effective infrared excess for the entire population of bright $z\sim7$ and $z\sim8$ we have targeted as part of multiple programs by dividing the mean SFR$_{IR}$ by the mean SFR$_{UV}$.  Using our bootstrap stacking results from section \ref{sec:stacking}, we can compute the median ratio and errors (68\%) on the obscured and UV luminous star formation rates to be 0.5$_{-0.3}^{+0.3}$ and 0.6$_{-0.3}^{+0.2}$, respectively.  This suggests that unobscured SFRs measured from the rest-$UV$ data should be corrected by factors of 1.5$_{-0.3}^{+0.3}$ (0.18$_{-0.10}^{+0.08}$ dex) for $z\sim7$ and 1.6$_{-0.3}^{+0.2}$ (0.20$_{-0.09}^{+0.05}$ dex) for $z\sim8$.

We can illustrate the impact these corrections have on the estimated SFR densities at $z\sim7$-8 from $UV$ bright galaxies based on the $UV$ LFs derived by Bouwens et al.\ (2015) and Stefanon et al.\ (2019) integrated down to $-$21.5 mag ($\sim$1.7 $L_{UV}^{*}$).  Applying these corrections to the UV luminosity density Bouwens et al.\ (2015) derive at $z\sim7$ and Stefanon et al.\ (2019) derive at $z\sim8$, we show the impact of these corrections on the computed SFR densities at $z\sim7$-8 in Figure~\ref{fig:sfrd}.  Our ALMA results suggest that $\sim$$\frac{1}{3}$ of the SFR density in $\gtrsim$1.7$L^*$ galaxies is obscured by dust. This is consistent with the 20-25\% fraction of the cosmic star formation rate density that  \citet{Zavala_2021} infer to be obscured at $z\sim6$-7, but it is worth emphasizing that the fraction we infer is for the $UV$-bright $z\sim7$ population and the fraction \citet{Zavala_2021} derive is with respect to the total.

Finally, we can combine the present constraints on the IR luminosities of sources with the $z\sim7$-8 rest-$UV$ luminosity function results of \citet{bowler2017_10.1093/mnras/stw3296} and \citet{Stefanon2019} to place constraints on the IR LF at $z>3$ such as discussed in \citet{Koprowski2017},  \citet{Casey_2018}, \citet{Gruppioni2020A&A...643A...8G}, and \citet{Zavala_2021}.  We find that the number density of galaxies with $L_{IR}>10^{11}-10^{12} L_{\odot}$/yr is $\gtrsim2\times10^{-6}$ Mpc$^{-3}$ dex$^{-1}$.  This limit is fairly consistent with the dust poor model described in \citet{Casey_2018}, being $\sim$0.3 dex lower.  This lower however excludes the model evolution of the IR luminosity function presented in \citet{Koprowski2017}. 

\section{Summary} \label{sec:conclusion}

We describe the analysis of sensitive ALMA continuum observations we have obtained over 15 bright Lyman-break galaxies at $z\sim7$-8.  The ALMA continuum observations we consider (25.7 hours in total on source) are drawn from five distinct ALMA programs (Table~\ref{tab:tab1}) and allow us to assess the contribution of obscured star formation to the build-up of stellar mass in these galaxies.    The  observations are sensitive enough to probe down to unobscured SFRs of $\sim$20 $M_{\odot}$/yr (3$\sigma$).  The bright Lyman-break galaxies we have targeted were drawn from a selection of the brightest $z\sim7$-8 galaxies identified over the UltraVISTA \citep[][S. Schouws et al.\ 2021, \textit{in prep}]{Stefanon2017,Stefanon2019} and CANDELS COSMOS field \citep{Smit_2015,Smit_2018Natur.553..178S}.

Out of the 15 $z\sim7$-8 galaxies we target, we detect six in the dust continuum, with inferred IR luminosities ranging from $2.7\times10^{11}$ $L_{\odot}$ to $1.1\times10^{12}$ $L_{\odot}$, with a median of $3.5\times10^{11}$ $L_{\odot}$.  This is equivalent to obscured SFRs of 25 to 101 $M_{\odot}$/yr (Figure~\ref{fig:postage_schouws}).  These dust detections more than double the number of normal star-forming galaxies at $z>6.5$ which are detected in the IR continuum from four \citep{Watson_2015_Natur.519..327W,Laporte_2017,Tamura_2019,Bowler_2018_10.1093/mnras/sty2368,Hashimoto_2019_10.1093/pasj/psz049} to 10 (Figure~\ref{fig:dustz}).  We find the spatial position of dust-continuum detections to be approximately cospatial with the rest-$UV$ light (median offset $<$0.2"), with UVISTA-Y-003 being a notable exception.  For UVISTA-Y-003, the dust-continuum detection is offset by $\sim$0.3$"$ from three distinct star-forming clumps seen in the rest-$UV$ (Figure~\ref{fig:postage_highres}).

Using the new dust-continuum detections, we can quantify the relative contribution that obscured and unobscured star formation provide to bright galaxies at $z\sim7$-8.  For the 6 bright galaxies in our sample which show ALMA continuum detections, we find that obscured SFR in these galaxies is comparable to the unobscured SFR in the median (Figure~\ref{fig:sfrirsfruv}).  

We also use our observations to look at the measured infrared excess IRX ($L_{IR}/L_{UV}$) in these galaxies and quantify how the infrared excess depends on the $UV$-continuum slope $\beta_{UV}$ and the stellar mass.  The infrared excess IRX is well known to correlate with both quantities, but there continues to be significant debate regarding the precise trend at $z>6$ with both $\beta_{UV}$ and stellar mass.  

Our IRX vs. $\beta_{UV}$ results show a significant dependence on what we assume for the dust temperature of $z\sim7$-8 galaxies as well as what we assume for the unreddened $UV$ slope $\beta_{UV,intr}$ for star-forming galaxies at $z\sim7$-8.  If we adopt a modified blackbody SED with a dust temperature of 50 K and an opacity index $\beta=1.6$ as our fiducial far-IR SED, we find current IRX vs. $\beta$ results are most consistent with an SMC dust curve if $\beta_{UV,intr}=-2.63$ and lie somewhere between a Calzetti and SMC dust curve if $\beta_{UV,intr}=-2.23$ (Figure~\ref{fig:IRX-beta}).  Adopting the same fiducial SEDs for interpreting IRX trends with stellar mass, we find results (Figure~\ref{fig:IRX-Mstar_2}) which are consistent with no evolution in IRX-stellar mass relation from $z\sim0$ \citep{Whitaker_2017}. 

Additionally, we use our new dust detections to estimate the total mass in dust in bright $z\sim7$-8 galaxies, for dust ejected from SN after reverse shock destruction. The dust masses we derive are dependent on the assumed dust temperature and range from $\sim 10^8M_{\odot}$ for low dust temperatures (T$\sim$30K) to $\sim10^7M_{\odot}$ for high temperatures (T$\sim$60K) (Figure~\ref{fig:mdust}). Combining these dust masses with the estimates for the stellar mass from SED fitting enable constraints on dust formation mechanisms (following \citet{michalowski2015A&A...577A..80M}. This shows that the our observed dust masses could be most likely be explained by dust production from SNe with low dust destruction (less than $\sim$90\%: see bottom panel of Figure~\ref{fig:mdust}).

The new measurements we make of the obscured SFRs in bright $z\sim7$-8 allow us to improve our estimates of the contribution very luminous bright galaxies make to the SFR density at $z\sim7$-8.  Incorporating the contribution from obscured star formation for both our bright $M_{UV,AB}<-21.5$ ($\gtrsim$1.7$L_{UV}^*$) $z\sim7$ and $z\sim8$ sample to the unobscured SFRs in these galaxies, we measure correction factors of 1.5$_{-0.3}^{+0.3}$ (0.18$_{-0.10}^{+0.08}$ dex) and 1.6$_{-0.3}^{+0.2}$ (0.20$_{-0.09}^{+0.05}$ dex), respectively.  These results clearly illustrate the important role that obscured star formation can play in the early stellar mass build-up of galaxies at $z\geq7$.  

Finally, we use these results to set a lower limit on the IR luminosity function of $z\sim7$-8 galaxies, finding $\gtrsim2\times10^{-6}$ Mpc$^{-3}$ dex$^{-1}$ for galaxies with  $L_{IR}>10^{11}-10^{12}$ $L_{\odot}$/yr.  This limit is most consistent with the dust poor model of \citet{Casey_2018}, being $\sim$0.3 dex lower.

In the future, we expect a significant increase in the number of dust detected galaxies in normal star-forming galaxies at $z\sim7$-8 from the ongoing ALMA observations from the REBELS ALMA Large program (2019.1.01634.L).  This should allow us to look at the build-up of dust in massive galaxies in much more detail, while allowing for a much more detailed look into how the measured IRX depends on both stellar mass and the $UV$ slope.

\acknowledgements
\subsection*{Acknowledgements}
We are greatly appreciative to our ALMA program coordinators Daniel Harsono and Carmen Toribio at ALLEGRO for support with our ALMA programs.  This paper makes use of the following ALMA data: ADS/JAO.ALMA\#2017.1.01217.S,\\ ADS/JAO.ALMA\#2017.1.00604.S,\\ ADS/JAO.ALMA\#2018.1.00236.S,\\ ADS/JAO.ALMA\#2018.1.00085.S\\ and ADS/JAO.ALMA\#2018.A.00022.S. ALMA is a partnership of ESO (representing its member states), NSF (USA) and NINS (Japan), together with NRC (Canada), MOST and ASIAA (Taiwan), and KASI (Republic of Korea), in cooperation with the Republic of Chile. The Joint ALMA Observatory is operated by ESO, AUI/NRAO and NAOJ.  SS and RB acknowledge support from  TOP grant TOP1.16.057 and a NOVA (Nederlandse Onderzoekschool Voor Astronomie) 5 grant. RS acknowledges support from an STFC Ernest Rutherford Fellowship (ST/S004831/1). JH acknowledges support of the VIDI research programme with project number 639.042.611, which is (partly) financed by the Netherlands Organisation for Scientific Research (NWO). SC acknowledges support from the European Research Council No. 74120 INTERSTELLAR. \\

\newpage
\appendix

\section{ALMA Observations of Bright $z\sim7$-8 Galaxies that are Undetected in the Dust Continuum}

\begin{figure*}[tbh]
\epsscale{1.0}
\plotone{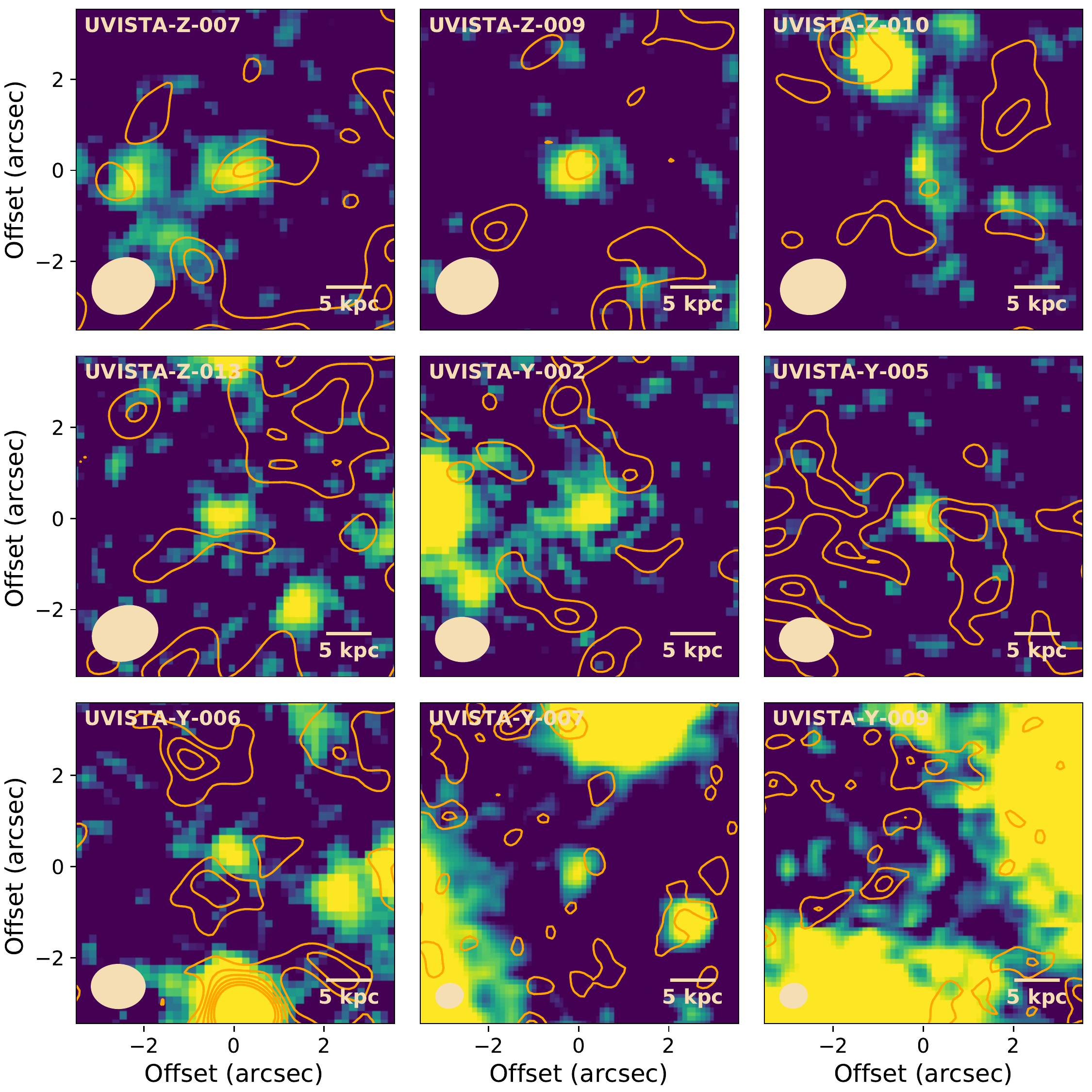}
\caption{Dust-continuum observations (1.2$\mu$m) overlaid on the available near-IR imaging observations from UltraVISTA (stacked J+H+K$_s$). Shown are the 9 $z\sim7$-8 sources that were not detected in the dust continuum. Contours are drawn at (1, 2, 3, 4, 5, 6)$\times\sigma$ where $\sigma\approx15\mu$Jy beam$^{-1}$. The synthesized beam for the dust-continuum observations is indicated in the lower-left corner of each panel. 
\label{fig:nondetections}}
\end{figure*}

Nine of the 15 bright $z\sim7$-8 galaxies (60\%) targeted by the ALMA programs considered here did not show $3\sigma$ detections in the dust continuum.  Figure~\ref{fig:nondetections} shows rest-$UV$ images of those nine sources along with (1, 2, 3, 4, 5)$\sigma$ contours showing the $IR$ dust continuum emission.  Interestingly enough, in the observations we obtain of UVISTA-Y-006, one foreground galaxy in our ALMA beam shows a clear dust-continuum detection (lowest row in Figure~\ref{fig:nondetections}, left panel).

\bibliography{references}



\end{document}